\documentclass{aa}  
\usepackage{dcolumn}
\usepackage{longtable}
\usepackage{graphicx,amsmath}
\usepackage{txfonts}
\newcommand{\Ha}{H$\alpha$}
\newcommand{\Hb}{H$\beta$}

\newcommand{\Nl}[3]{#1\,{\sc #2}\,$\lambda{#3}$}
\makeatletter
 \def\hlinewd#1{%
   \noalign{\ifnum0=`}\fi\hrule \@height #1 \futurelet
    \reserved@a\@xhline}
\makeatother
\newcommand{\htopline}{\hlinewd{.8pt}}
\newcommand{\hmidline}{\hlinewd{.2pt}}
\newcommand{\hbotline}{\htopline}
\newcommand{\mcc}[1]{\multicolumn{1}{c}{#1}}

\newcommand{\mcr}[1]{\multicolumn{1}{r}{#1}}
\newcommand{\kms}{km\,s$^{-1}$}

\begin{document} 

    \title{Broad-line region structure and line profile variations
 in the changing look AGN HE\,1136-2304
\thanks{Based on observations obtained with the Southern African
 Large Telescope.}
}

   \author{W. Kollatschny \inst{1}, 
           M. W. Ochmann \inst{1},
           M. Zetzl \inst{1}, 
           M. Haas \inst{2},
           D. Chelouche \inst{3}, 
           S. Kaspi \inst{4}, 
           F. Pozo Nu\~nez\inst{3},
           D. Grupe \inst{5} 
            }

   \institute{Institut f\"ur Astrophysik, Universit\"at G\"ottingen,
              Friedrich-Hund Platz 1, D-37077 G\"ottingen, Germany\\
              \email{wkollat@astro.physik.uni-goettingen.de}
          \and
          Astronomisches Institut, Ruhr-Universit\"at Bochum,
               Universit\"atsstrasse 150, 44801 Bochum, Germany
         \and
   Physics Department and the Haifa Research Center for Theoretical Physics and
   Astrophysics, University of Haifa, Haifa 3498838, Israel
         \and  
          School of Physics \& Astronomy and the Wise Observatory,
   The Raymond and Beverly Sackler Faculty of Exact Sciences
   Tel-Aviv University, Tel-Aviv 69978, Israel
         \and  
          Department of Earth and Space Sciences, Morehead State University, Morehead, KY 40351, USA
}

   \date{Received 27 June 2018; Accepted 7 August 2018}
   \authorrunning{Kollatschny et al.}
   \titlerunning{HE\,1136-2304}

 
  \abstract
   {}
   {A strong X-ray outburst was detected in HE\,1136-2304 in 2014.
Accompanying optical spectra revealed that the spectral type has changed
from a nearly Seyfert 2 type (1.95), classified by 
spectra taken 10 and 20 years ago, 
to a Seyfert 1.5 in our most recent observations. We seek to investigate
a detailed spectroscopic campaign on the
spectroscopic properties  and spectral variability
behavior of this changing look AGN and compare this to other variable
Seyfert galaxies.}
   {We carried out a detailed spectroscopic variability campaign of
   HE\,1136-2304 with the 10 m Southern African Large Telescope (SALT)
  between
   2014 December and 2015 July.}  
   {The broad-line region (BLR) of HE\,1136-2304 is stratified with respect
 to the distance of
the line-emitting regions. The integrated emission line intensities
 of H$\alpha$, H$\beta$,
\ion{He}{i}\,$\lambda 5876$, and \ion{He}{ii}\,$\lambda 4686$ originate at
distances of $15.0^{+4.2}_{-3.8}$, $7.5^{+4.6}_{-5.7}$, $7.3^{+2.8}_{-4.4}$, and
$3.0^{+5.3}_{-3.7}$ light days with respect to the optical continuum
at 4570\,\AA{}.
The variability amplitudes of the integrated emission lines
are a function of distance to the ionizing continuum source as well.
We derived a central black hole mass of
$3.8 \pm 3.1 \times 10^{7} M_{\odot}$ based on the linewidths
 and distances of the BLR.
The outer line wings of all BLR  lines respond much faster
to continuum variations 
indicating a Keplerian
disk component for the BLR. The response in
the outer wings is about two light days shorter than the response
of the adjacent continuum
flux with respect to the ionizing continuum flux. 
The vertical BLR structure in HE\,1136-2304
confirms a general trend that
 the emission lines of narrow line active galactic nuclei (AGNs) originate
 at larger
distances from the midplane in comparison to AGNs showing broader
emission lines.
Otherwise, the variability behavior of this changing look AGN 
is similar to that of other AGN.}
{}
\keywords {Galaxies: active --
                Galaxies: Seyfert  --
                Galaxies: nuclei  --
                Galaxies: individual: HE\,1136-2304 --   
                (Galaxies:) quasars: emission lines 
               }

   \maketitle
%

\section{Introduction}

About a dozen Seyfert galaxies are known to have significantly changed their
optical spectral type: for example, NGC\,3515 (Collin-Souffrin
et al.\citealt{souffrin73}),
 NGC\,4151 (Penston \& Perez\citealt{penston84}), Fairall\,9 
 (Kollatschny et al.\citealt{kollatschny85}), NGC\,2617 (Shappee
 et al.\citealt{shappee14}), Mrk\,590 (Denney at al.\citealt{denney14}), 
and references therein. Further recent findings are based on spectral
variations detected by means of the Sloan Digital Sky Survey 
(e.g., Komossa et al.\citealt{komossa08}; LaMassa et al.\citealt{lamassa15};
Runnoe et al.\citealt{runnoe16}; MacLeod et al.\citealt{macleod16}).
These galaxies are considered to be changing look
active galactic nuclei (AGNs). 
However, most of these findings are based on only a few optical spectra. 

HE\,1136-2304 ($\alpha_{2000}$ = 11h 38m 51.1s,
$\delta_{2000}$ = $-23^{\circ}$ 21$^{'}$ 36$^{''}$)
was classified as a changing look AGN based on spectroscopy performed after a 
strong increase in the X-ray flux was detected by  
XMM-Newton in 2014
in comparison to an upper limit based on the ROSAT All-Sky Survey
taken in 1990
(Parker et al.\citealt{parker16}). The increase in the X-ray flux came 
with an increase in the optical continuum flux and with a change
of the Seyfert type. HE\,1136-2304
was of Seyfert 2/1.95 type in early spectra taken in 1993 and 2002. 
However, its spectral type changed to a Seyfert 1.5 type
in 2014 (Zetzl et al.\citealt{zetzl18}, Paper 1). This notation 
of Seyfert subclasses was introduced by Osterbrock \cite{osterbrock81}.

Long-term and detailed optical variability
studies exist for many AGN such as
NGC\,5548 (Peterson et al.\citealt{peterson02}; Pei et al.\citealt{pei17}
and references therein), 3C120 (Peterson et al.\citealt{peterson98};
Kollatschny et al.\citealt{kollatschny00};  Grier at al.\citealt{grier13}),
NGC\,7603 (Kollatschny et al.\citealt{kollatschny00}),
and 3C\,390.3  (Shapovalova et al.\citealt{shapovalova10}).
Corresponding detailed
follow-up studies have not yet been reported for 
the type of changing look AGN mentioned above.

We carried out a detailed spectroscopic and photometric variability
study of HE\,1136-2304 between  2014 and 2015 
after the detection of the strong outburst in 2014.
We presented the optical, UV, and X-ray continuum variations
of HE\,1136-2304 from
2014 to 2017 in a separate paper (Paper 1).
We  verified strong continuum variations in the X-ray, UV, and
optical continua.
We showed that the variability amplitude decreased with increasing wavelength.
The amplitude in the optical varied by a factor of three after
correcting for the host galaxy contribution.
 No systematic trends were
 found with regards to the variability behavior following
the outburst in 2014. A general decrease in flux would have been expected
for a tidal disruption event.
The Seyfert type did not change between 2014 and 2017 despite strong
continuum variations.
We describe the results of the spectroscopic variability campaign
taken with the 10 m Southern African Large Telescope (SALT) 
for the years 2014 to 2015.

Throughout this paper, we assume $\Lambda$CDM cosmology with
a Hubble constant of H$_0$~=~70~\kms Mpc$^{-1}$, $\Omega_{\text{M}}$=0.27,
and $\Omega_{\Lambda}$=0.73. Following the cosmological calculator
by Wright et al.
(\citealt{wright06}) this results in a luminosity distance of 118 Mpc.

\section{Observations and data reduction}

In addition to our first spectrum obtained on 2014 July 7 
(Parker et al.\citealt{parker16}), we took optical
spectra of the Seyfert galaxy HE\,1136-2304 with the SALT
 at 17 epochs between 2014 December 25
and 2015 July 13 .
 The log of our spectroscopic observations is given in
Tab.~\ref{log_of_SALT_obs}.   
\begin{table}
\tabcolsep+7mm
\caption{Log of spectroscopic observations of HE\,1136-2304 with SALT}
\centering
\begin{tabular}{ccc}
\hline 
\noalign{\smallskip}
Julian Date & UT Date & Exp. time \\
2\,400\,000+&         &  [sec.]   \\
\hline 
56846.248       &       2014-07-07      &      1200     \\
57016.559       &       2014-12-25      &       985     \\
57070.399       &       2015-02-16      &       985     \\
57082.362       &       2015-02-28      &       985     \\
57088.594       &       2015-03-07      &       985     \\
57100.539       &       2015-03-19      &       985     \\
57112.285       &       2015-03-30      &       985     \\
57121.256       &       2015-04-08      &       985     \\
57131.243       &       2015-04-18      &      1230     \\
57167.359       &       2015-05-24      &      1144     \\
57171.364       &       2015-05-28      &      1144     \\
57182.330       &       2015-06-08      &      1144     \\
57187.319       &       2015-06-13      &      1144     \\
57192.308       &       2015-06-18      &      1144     \\
57196.295       &       2015-06-22      &      1144     \\
57201.271       &       2015-06-27      &      1144     \\
57206.265       &       2015-07-02      &      1144     \\
57217.227       &       2015-07-13      &      1144     \\
\hline 
\vspace{-.7cm}
\end{tabular}
\label{log_of_SALT_obs}
\end{table}
The spectra taken of HE\,1136-2304 between 2015 February and 2015 July
had a mean interval of nine days. For some epochs  spectra were acquired 
at shorter intervals.

All spectroscopic observations were taken with identical instrumental setups.
We used the Robert Stobie Spectrograph
attached to the SALT using the PG0900 grating.
The slit width was fixed to
2\arcsec\hspace*{-1ex}.\hspace*{0.3ex}0 projected on the
sky at an optimized position angle to minimize differential refraction.
Furthermore, all observations were taken at the same airmass
owing to the particular design of the SALT.
The spectra were taken with exposure times of 
16 to 20 minutes.

Typical seeing values were 1 to 2 arcsec.
We covered a wavelength range from
4355 \AA \ to 7230~\AA,\  which corresponds to a rest-frame wavelength range of 
4240 \AA\  to 7040~\AA{}. The  spectral resolution was 6.5~\AA{}.
 There are two gaps in the spectrum caused by the gaps between the three CCDs:
one between the blue and the central CCD chip and one between the
central and red CCD chip covering the wavelengths in the ranges
5206--5263~\AA\  and 6254--6309~\AA\ (5069--5124~\AA\ and 6089--6142~\AA\
in the rest frame){.}
All spectra shown in this work were shifted to the rest frame
of HE 1136-2604.

In addition to the galaxy spectrum, necessary flat-field and
Xe arc frames were also observed, as well
as spectrophotometric standard stars for flux calibration 
(LTT3218, LTT7379, and EG274).
Flat-field frames
were used to correct for differences in sensitivity both
between detector pixels and across the field.
The spatial
resolution per binned pixel was 0\arcsec\hspace*{-1ex}.\hspace*{0.3ex}2534
for our SALT spectrum.
We extracted eight columns from the object spectrum 
corresponding to 2\arcsec\hspace*{-1ex}.\hspace*{0.3ex}03.

We reduced the spectra (bias subtraction, cosmic ray correction,
flat-field correction, 2D-wavelength calibration, night sky subtraction, and
flux calibration) in a homogeneous way with the
Image Reduction and Analysis Facility (IRAF) reduction
packages (e.g., Kollatschny et al.,\citealt{kollatschny01}). 
Great care was taken to ensure high-quality intensity and wavelength
calibrations to keep the intrinsic measurement errors very low
(Kollatschny et al.,\citealt{kollatschny01,kollatschny03,kollatschny10}).
The spectra of HE1136-2304 and the
 calibration star spectra were not always taken
 under photometric conditions. 
Therefore,
all spectra were calibrated to the same absolute
[\ion{O}{iii}]\,$\lambda$5007 flux of
$1.75 \times 10^{-13} \text{erg\,s}^{-1}\,\text{cm}^{-2}$
 (Reimers et al.\citealt{reimers96}).
The flux of the narrow emission line [\ion{O}{iii}]\,$\lambda$5007
is considered to be constant on timescales of years.
The accuracy of the [\ion{O}{iii}]\,$\lambda$5007 flux calibration
 was tested for all forbidden emission lines in the spectra.
 We calculated difference spectra for all epochs
 with respect to the mean spectrum of our variability campaign.
Corrections for both small spectral shifts ($<$ 0.5 \AA )
 and small scaling factors were executed
 by minimizing the residuals of the narrow emission lines in the
 difference spectra. A relative flux  accuracy on the order
of 1\% was achieved for most of the spectra.

\section{Results}
\subsection{Continuum and spectral line variations}

\begin{figure*}
\centering
\includegraphics[width=10cm,angle=-90]{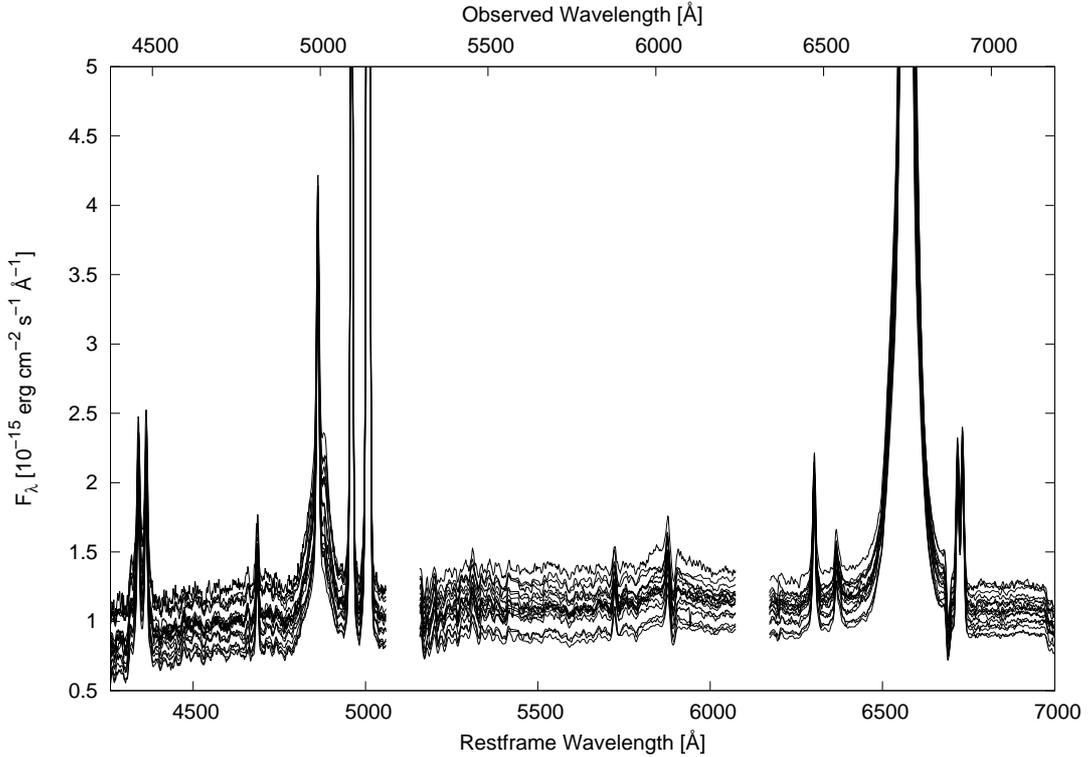}
       \vspace*{-2mm} 
  \caption{Optical spectra of HE\,1136-2304 taken with the SALT telescope
   for our variability campaign from December 2014 until July 2015.
}
   \label{spec_all_intercali.ps}
\end{figure*}

\begin{figure*}
 \centering
   \includegraphics[width=10cm,angle=-90]{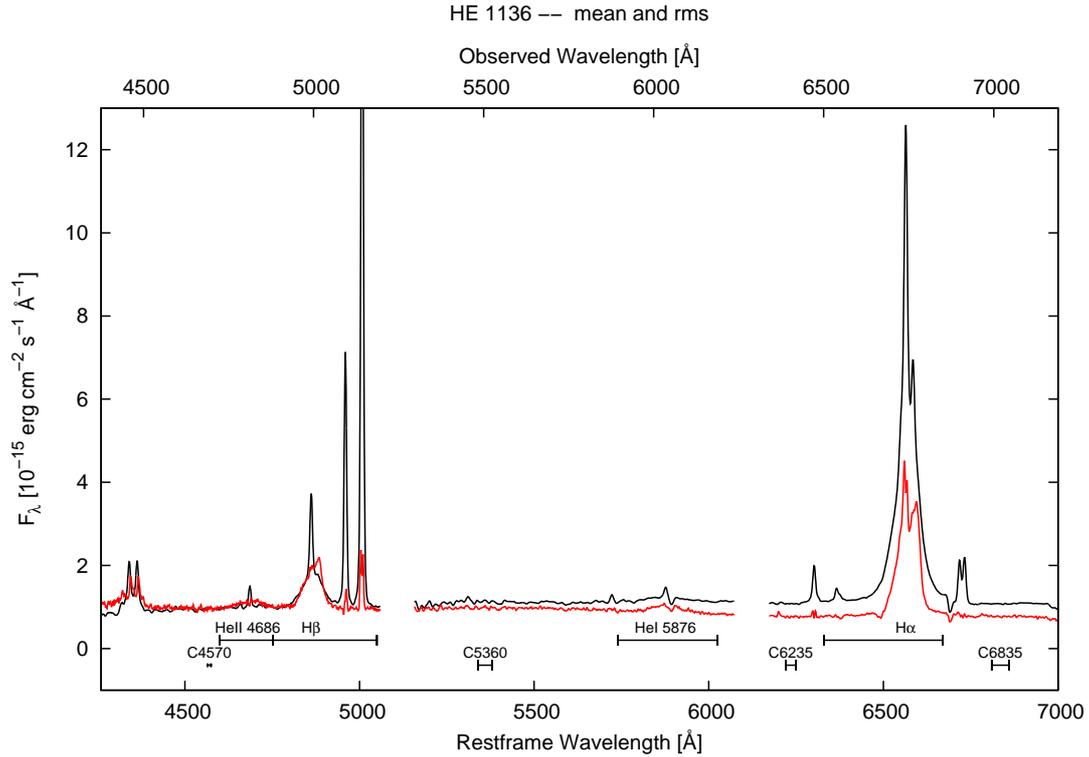}
      \caption{Integrated mean (black) and rms (red) spectra
               for our variability campaign of HE\,1136-2304.
               The rms spectrum has been scaled by a factor
               of 6.9 to enhance  weak line structures.
                }
       \vspace*{-3mm} 
         \label{spec_meanrms_intercali.ps}
   \end{figure*}

We present all  final reduced optical spectra of HE\,1136-2304
taken during the 2014/2015 variability campaign 
in Fig.~\ref{spec_all_intercali.ps}.
These results clearly show variations in the
continuum intensities.
The mean and rms spectra of HE\,1136-2304 are shown in
Fig.~\ref{spec_meanrms_intercali.ps}.
The rms spectrum presents the variable part of the emission lines.
The spectrum was scaled by a factor
of 6.9 to allow for a better comparison with the mean spectrum
and for enhancing weaker line structures. We note that all wavelengths
referred to in this section are rest-frame wavelengths.

%
%
%
%
%

The integration limits of the broad emission lines and continuum regions
 are given at the bottom of the spectra.
To select the continuum regions,
we inspected the mean and rms spectra for regions that are
free of both strong emission and absorption lines. The final wavelength
ranges used for our continuum flux measurements are given in
Table~\ref{cont_boundaries}.
A continuum region at 5100 \AA\ is often used
in studies of the variable continuum flux in AGN.
Normally, this region is free of strong emission lines and close
to the [\ion{O}{iii}]\,$\lambda$5007 flux calibration line.
However, in our case this region falls in the gap between the blue and
central CCD chip. Therefore we set a nearby continuum range
at 5360 \AA{}. In addition to this continuum range, we determined
the continuum intensities 
at three additional ranges (at 4570, 6235,
and 6835 \AA; see Fig.~\ref{spec_meanrms_intercali.ps}
and {Tab.~\ref{cont_boundaries}}).
We used these continuum regions for creating pseudo-continua
below the variable broad emission lines. We neglected a possible
contribution of FeII blends to the continuum flux based on our
mean and rms spectra. 
\begin{table}
\centering
\tabcolsep+3.5mm
\caption{Rest-frame continuum boundaries and line integration limits.}
\begin{tabular}{lcc}
\hline 
\noalign{\smallskip}
Cont./Line                   & Wavelength range & Pseudo-continuum \\
\noalign{\smallskip}
(1)                           & (2)                             & (3) \\
\noalign{\smallskip}
\hline 
\noalign{\smallskip}
Cont.~4570                    & 4565\,\AA{} -- 4575\,\AA{}     & \\
Cont.~5360                    & 5340\,\AA{} -- 5380\,\AA{}     & \\
Cont.~6235                    & 6220\,\AA{} -- 6250\,\AA{}     & \\
Cont.~6835                    & 6810\,\AA{} -- 6860\,\AA{}     & \\
\ion{He}{ii}\,$\lambda 4686$  & 4600\,\AA{} -- 4752\,\AA{}     & 4570\,\AA{} -- 5360\,\AA{} \\
H$\beta$                      & 4752\,\AA{} -- 5050\,\AA{}     & 4570\,\AA{} -- 5360\,\AA{} \\
\ion{He}{i}\,$\lambda 5876$   & 5740\,\AA{} -- 6025\,\AA{}     & 5700\,\AA{} -- 6060\,\AA{} \\
H$\alpha$                     & 6330\,\AA{} -- 6670\,\AA{}     & 6235\,\AA{} -- 6835\,\AA{} \\
\noalign{\smallskip}
\hline \\
\vspace{-1.1cm}
\end{tabular}
\label{cont_boundaries}
\end{table}
%

We integrated the
broad Balmer and Helium emission-line intensities 
between the wavelength boundaries given in Tab.~\ref{cont_boundaries}.
Before integrating each emission line flux,
we subtracted a linear pseudo-continuum 
defined by the boundaries given in Tab.~\ref{cont_boundaries} (Col.\,3).
We did not consider the H$\gamma$ line in our studies as it was not possible
to determine a reliable continuum at the blue side of this line. 
The results of the continuum and line intensity measurements are given in 
Tab.~\ref{cont_integline_intens}. Additionally, we present in this table
the flux values obtained for our first spectrum taken on 2014 July 7.
\begin{table*}
\tabcolsep4.7mm
\newcolumntype{d}{D{.}{.}{-2}}
\caption{Continuum and integrated broad line fluxes for different epochs.}
\begin{tabular}{ccrrrrr}
\noalign{\smallskip}
\hline 
\noalign{\smallskip}
Julian Date &   \mcr{Cont.~4570\,\AA}& \mcr{Cont.~5360\,\AA}    & \mcc{H$\alpha$}     &   \mcc{H$\beta$}    &  \mcc{HeI}     &  \mcc{HeII}    \\
2\,400\,000+\\
(1) & (2) & \mcc{(3)} & \mcc{(4)} & \mcc{(5)} & \mcc{(6)} & \mcc{(7)}\\
\noalign{\smallskip}
\hline
\noalign{\smallskip}
56846.248      &               1.336   $\pm$                   0.011   &       1.403   $\pm$           0.030   &               514.1   $\pm$                   15.5    &               145.2   $\pm$                   4.4     &                                                                       29.5            $\pm$                   0.9     &               17.2    $\pm$                   0.6     \\
57016.559       &               0.829   $\pm$                   0.003   &       0.889   $\pm$           0.019   &               479.0   $\pm$                   14.4    &               71.6    $\pm$                   3.6     &                                                               
        16.1            $\pm$                   0.5     &               4.3     $\pm$                   0.3     \\
57070.399       &               0.909   $\pm$                   0.012   &       0.927   $\pm$           0.021   &               444.9   $\pm$                   13.4    &               83.4    $\pm$                   2.6     &                                                               
        13.8            $\pm$                   0.5     &               8.1     $\pm$                   0.3     \\
57082.362       &               1.161   $\pm$                   0.019   &       1.191   $\pm$           0.025   &               451.8   $\pm$                   13.6    &               97.4    $\pm$                   3.0     &                                                               
        18.3            $\pm$                   0.6     &               6.3     $\pm$                   0.2     \\
57088.594       &               1.257   $\pm$                   0.028   &       1.199   $\pm$           0.032   &               508.3   $\pm$                   15.3    &               90.9    $\pm$                   2.8     &                                                               
        23.4            $\pm$                   0.8     &               9.7     $\pm$                   0.3     \\
57100.539       &               1.139   $\pm$                   0.003   &       1.162   $\pm$           0.021   &               547.7   $\pm$                   16.5    &               107.0   $\pm$                   3.3     &                                                               
        25.3            $\pm$                   0.8     &               11.0    $\pm$                   0.4     \\
57112.285       &               1.023   $\pm$                   0.015   &       1.043   $\pm$           0.025   &               534.6   $\pm$                   16.1    &               94.3    $\pm$                   2.9     &                                                               
        18.9            $\pm$                   0.6     &               6.9     $\pm$                   0.3     \\
57121.256       &               1.062   $\pm$                   0.009   &       1.112   $\pm$           0.020   &               515.4   $\pm$                   15.5    &               89.1    $\pm$                   2.7     &                                                               
        15.4            $\pm$                   0.5     &               5.0     $\pm$                   0.2     \\
57131.243       &               1.174   $\pm$                   0.010   &       1.196   $\pm$           0.026   &               526.2   $\pm$                   15.8    &               93.8    $\pm$                   2.9     &                                                               
        20.5            $\pm$                   0.7     &               10.0    $\pm$                   0.4     \\
57167.359       &               1.011   $\pm$                   0.016   &       1.068   $\pm$           0.024   &               498.7   $\pm$                   15.0    &               81.2    $\pm$                   4.1     &                                                               
        16.5            $\pm$                   0.5     &               6.1     $\pm$                   0.4     \\
57171.364       &               0.990   $\pm$                   0.010   &       1.013   $\pm$           0.027   &               513.8   $\pm$                   15.5    &               84.5    $\pm$                   2.6     &                                                               
        18.3            $\pm$                   0.6     &               8.1     $\pm$                   0.3     \\
57182.330       &               0.825   $\pm$                   0.008   &       0.875   $\pm$           0.020   &               473.4   $\pm$                   14.3    &               76.9    $\pm$                   2.4     &                                                               
        14.3            $\pm$                   0.5     &               3.0     $\pm$                   0.1     \\
57187.319       &               0.782   $\pm$                   0.015   &       0.864   $\pm$           0.014   &               488.1   $\pm$                   14.7    &               70.3    $\pm$                   2.2     &                                                               
        15.2            $\pm$                   0.5     &               1.5     $\pm$                   0.1     \\
57192.308       &               0.907   $\pm$                   0.009   &       0.969   $\pm$           0.024   &               484.7   $\pm$                   14.6    &               67.2    $\pm$                   3.4     &                                                               
        14.4            $\pm$                   0.5     &               5.5     $\pm$                   0.3     \\
57196.295       &               0.915   $\pm$                   0.011   &       0.966   $\pm$           0.023   &               436.0   $\pm$                   13.1    &               59.3    $\pm$                   1.8     &                                                               
        9.9             $\pm$                   0.3     &               2.2     $\pm$                   0.1     \\
57201.271       &               0.800   $\pm$                   0.008   &       0.868   $\pm$           0.022   &               444.3   $\pm$                   13.4    &               60.0    $\pm$                   1.9     &                                                               
        13.3            $\pm$                   0.5     &               1.8     $\pm$                   0.1     \\
57206.265       &               0.994   $\pm$                   0.012   &       1.067   $\pm$           0.023   &               441.1   $\pm$                   13.3    &               68.9    $\pm$                   3.5     &                                                               
        14.7            $\pm$                   0.5     &               6.9     $\pm$                   0.3     \\
57217.227       &               1.004   $\pm$                   0.024   &       1.091   $\pm$           0.032   &               452.0   $\pm$                   13.6    &               66.4    $\pm$                   2.0     &                                                               
        14.1            $\pm$                   0.5     &               2.0     $\pm$                   0.1     \\
%
\noalign{\smallskip}
\hline 
\noalign{\smallskip}
\end{tabular}
\tablefoot{
Continuum fluxes (2) - (3) in units of 10$^{-15}$\,erg\,s$^{-1}$\,cm$^{-2}$\,\AA$^{-1}$.\\
Line fluxes (3) - (7) in units 10$^{-15}$\,erg\,s$^{-1}$\,cm$^{-2}$.
}
\label{cont_integline_intens}
\end{table*}

\begin{figure*}
\centering
\vspace{-3mm}
 \includegraphics[width=16.cm,angle=0]{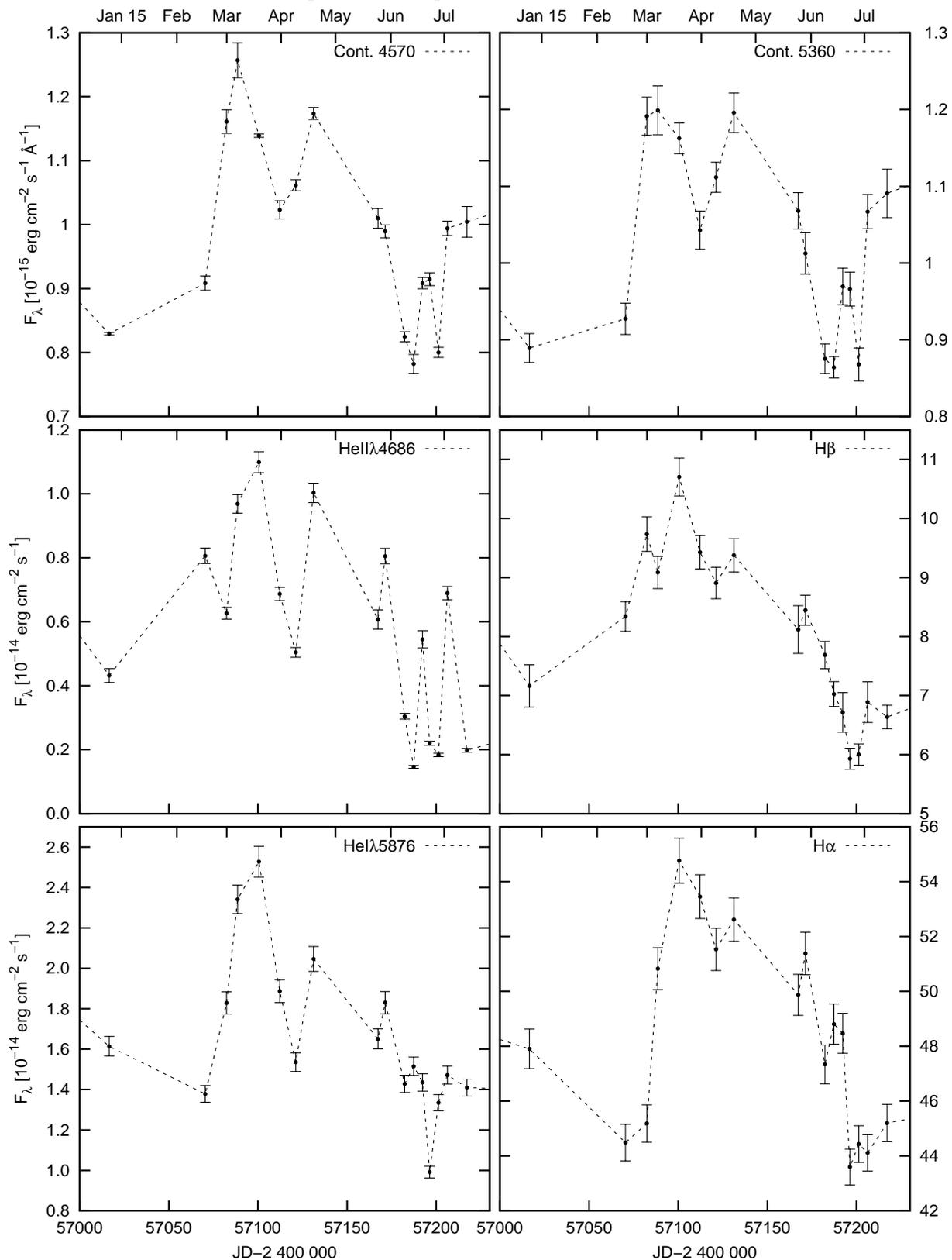}
  \caption{Light curves of the continuum fluxes at 4570\,\AA\
  and 5360\,\AA\
 (in units of 10$^{-15}$ erg cm$^{-2}$ s$^{-1}$\,\AA$^{-1}$) as well as of
the integrated emission line fluxes of H$\alpha$, H$\beta$,
 \ion{He}{ii}\,$\lambda 4686$, and \ion{He}{i}\,$\lambda 5876$
   (in units of  10$^{-14}$ erg cm$^{-2}$ s$^{-1}$) for
 our variability campaign from 2014 December until 2015 July.}
  \label{LC_SALT_2015_ochm.ps}
\end{figure*}

We present the light curves of the continuum fluxes at 4570\,\AA\
and 5360\,\AA\ as well as those of the integrated emission line fluxes
of H$\alpha$, H$\beta$,
 \ion{He}{ii}\,$\lambda 4686$, and
\ion{He}{i}\,$\lambda 5876$  in 
Fig.~\ref{LC_SALT_2015_ochm.ps}.
Some statistics of the emission line intensity and continuum variations
are given in Tab.~\ref{variab_statistics}.
We indicate the minimum and maximum fluxes F$_\text{min}$ and F$_\text{max}$,
peak-to-peak amplitudes R$_\text{max}$ = F$_\text{max}$/F$_\text{min}$,
the mean flux
over the period of observations $<$F$>$, standard deviation $\sigma_F$,
 and fractional variation F$_{\text{var}}$ (see Paper 1).
\begin{table}
\centering
\tabcolsep+2mm
\caption{Variability statistics based on the SALT data in units of
 10$^{-15}$\,erg\,s$^{-1}$\,cm$^{-2}$\,\AA$^{-1}$ for the continuum
 as well as in units of 
 10$^{-15}$\,erg\,s$^{-1}$\,cm$^{-2}$ for the emission lines.
}
\begin{tabular}{lccccccc}
\hline 
\noalign{\smallskip}
Cont./Line & F$_\text{min}$ & F$_\text{max}$ & R$_\text{max}$ & $<$F$>$ & $\sigma_\text{F}$ & F$_\text{var
}$ \\
\noalign{\smallskip}
\hline 
\noalign{\smallskip}
(1) & (2) & (3) & (4) & (5) & (6) & (7) \\ 
\noalign{\smallskip}
Cont.~4570 & 0.78 & 1.26 & 1.61 & 0.99 & 0.14 & 0.141 \\ 
Cont.~5360 & 0.86 & 1.20 & 1.39 & 1.03 & 0.12 & 0.114 \\ 
Cont.~6235& 0.88 & 1.29 & 1.46 & 1.08 & 0.11 & 0.102 \\ 
Cont.~6835 & 0.89 & 1.26& 1.41 & 1.08 & 0.11 & 0.103 \\
H$\alpha$ & 436. & 547.7 & 1.26 & 484.7 & 35.81 & 0.072 \\ 
H$\beta$  &  59.3 & 107.0 & 1.80 & 80.1 & 14.00 & 0.171 \\ 
\ion{He}{i}\,$\lambda 5876$ &  9.9 & 25.3 & 2.55 & 16.6 & 3.83 & 0.229 \\ 
\ion{He}{ii}\,$\lambda 4686$ & 1.5 & 11.0 & 7.52 & 5.78 & 3.01 & 0.520 \\
\hline 
\noalign{\smallskip}
\end{tabular}
\label{variab_statistics}
\end{table}

\begin{figure}
\centering
\includegraphics[width=6.5cm,angle=-90]{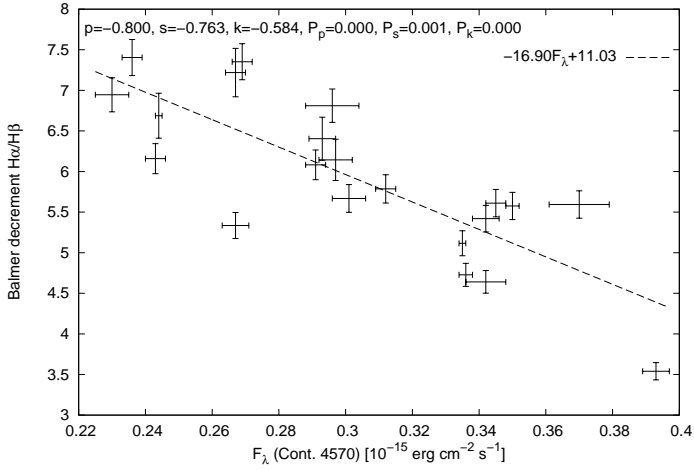}
\caption{Balmer decrement \Ha{}/\Hb{} of the broad line components vs. 
the continuum intensity at 4570 \AA{}.
The dashed line on the graph represents the linear regression.
}
\label{balmerdek_vs_cont4570.ps}
\end{figure}

\begin{figure}
\centering
\includegraphics[width=6.5cm,angle=-90]{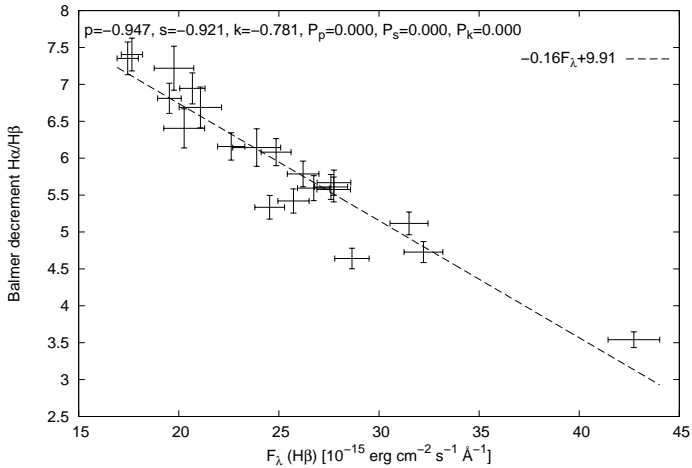}
\caption{Balmer decrement \Ha{}/\Hb{} of the broad line components vs. 
broad line \Hb{} intensity.
The dashed line on the graph represents the linear regression.
}
\label{balmerdek_vs_hb.ps}
\end{figure}

We calculated the Balmer decrement \Ha{}/\Hb{} of the broad components
after subtraction of the narrow components for the individual epochs.
The results are shown in Fig.~\ref{balmerdek_vs_cont4570.ps}
as a function of the continuum intensity at 4570 \AA{}.
Fig.~\ref{balmerdek_vs_hb.ps} gives the Balmer decrement
as a function of the \Hb{} line intensity.
The Balmer decrement \Ha{}/\Hb{} of the broad components
takes values between 3.5 and 7.5. On the other hand, we measure
a constant value of 2.81
for the Balmer decrement \Ha{}/\Hb{} 
of the narrow components.
There is a clear trend that the Balmer decrement \Ha{}/\Hb{}
 of the broad component decreases with increasing luminosity.


\subsection{Mean and rms line profiles}

\begin{figure}[t]
 \includegraphics[height=9.cm,angle=-90]{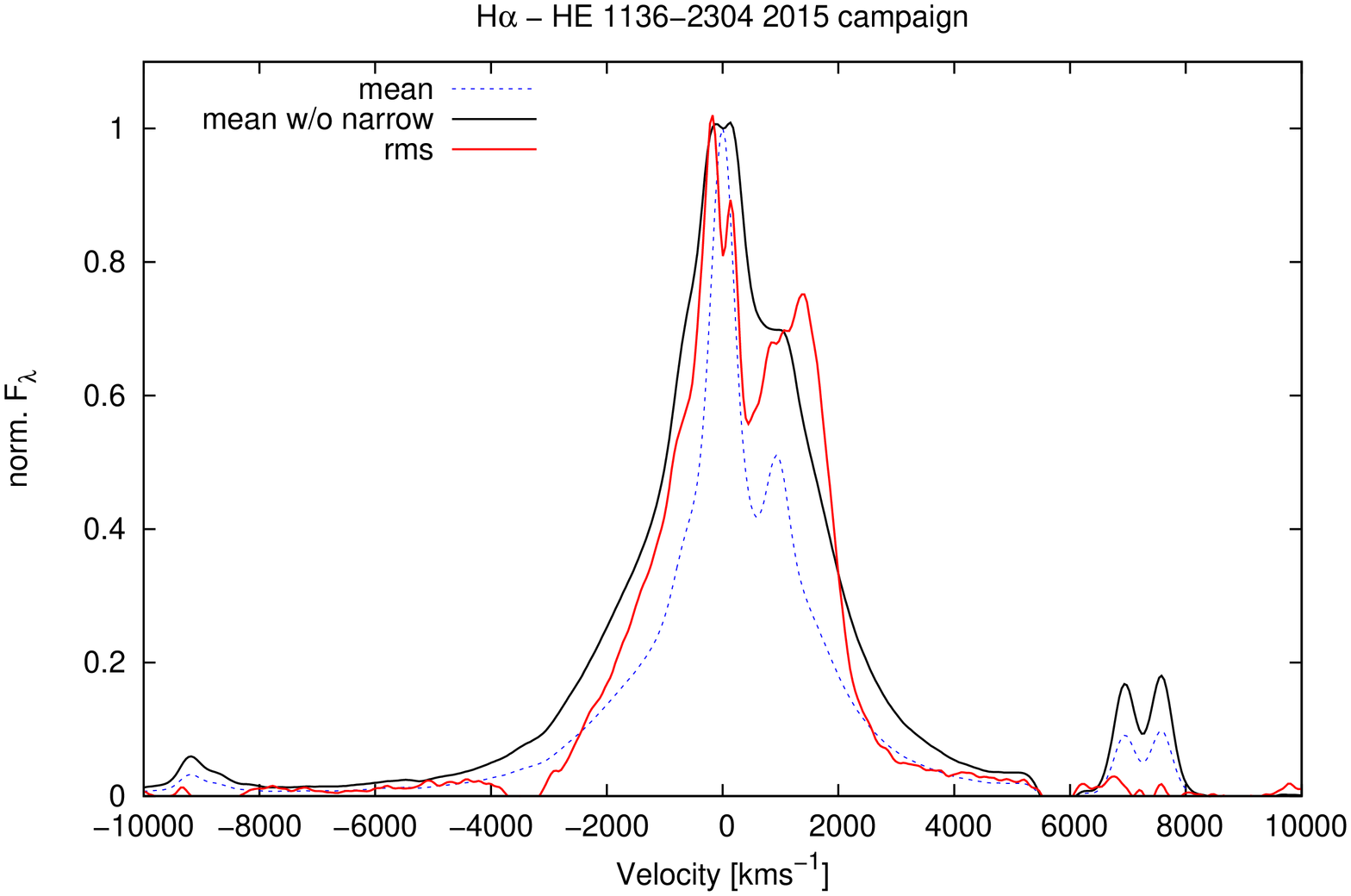}
  \caption{Normalized mean (blue), mean without a narrow component (black),
 and rms (red) line
 profiles of H$\alpha$
   in velocity space.}
  \label{velo_meanrms_ha.ps}
%
  \includegraphics[height=9.cm,angle=-90]{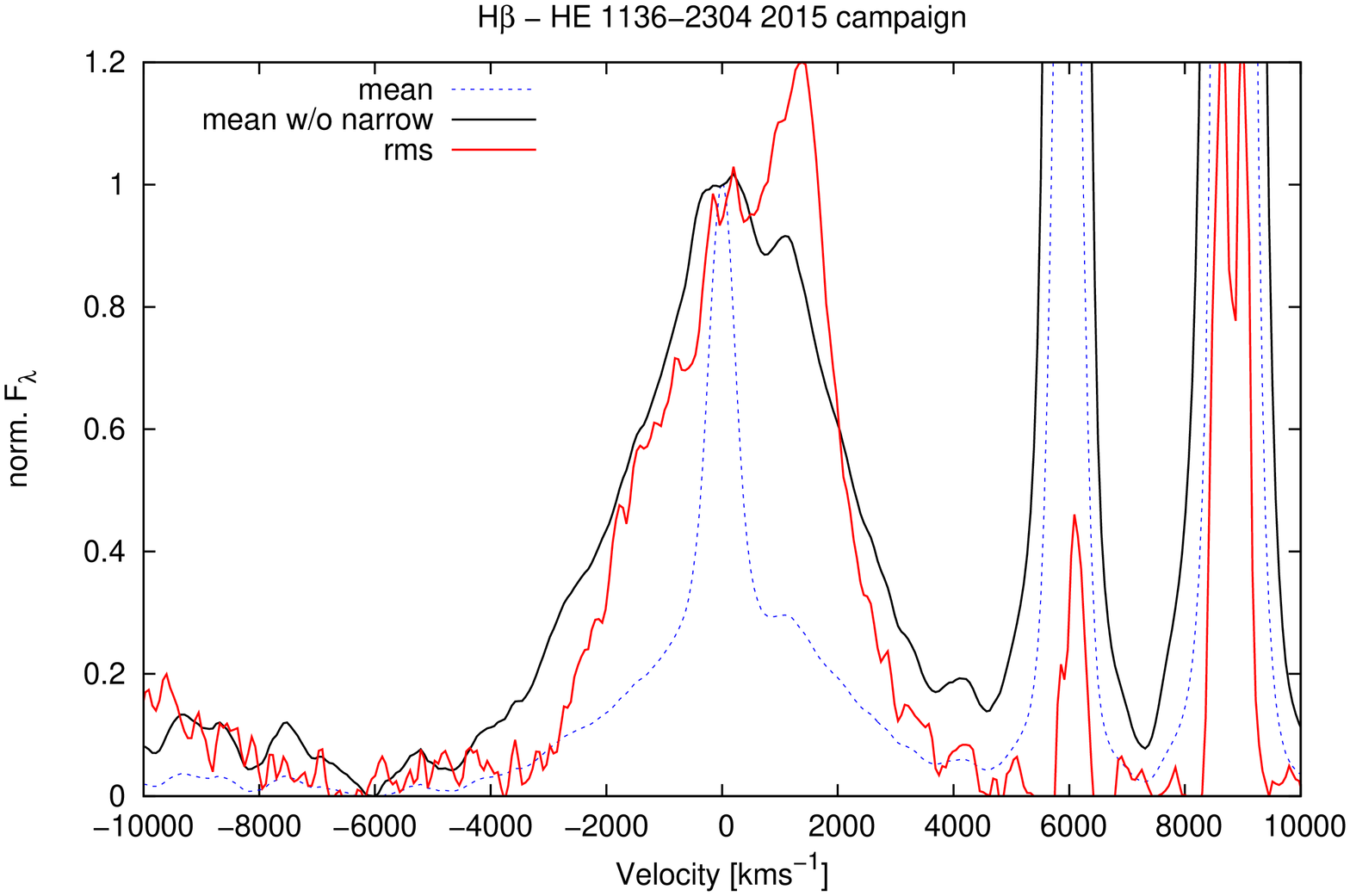}
  \caption{Normalized mean (blue), mean without a narrow component (black),
  and rms (red)
 line profiles of H$\beta$ in velocity space.}
  \label{velo_meanrms_hb.ps}
\end{figure}

\begin{figure}
  \includegraphics[height=9.cm,angle=-90]{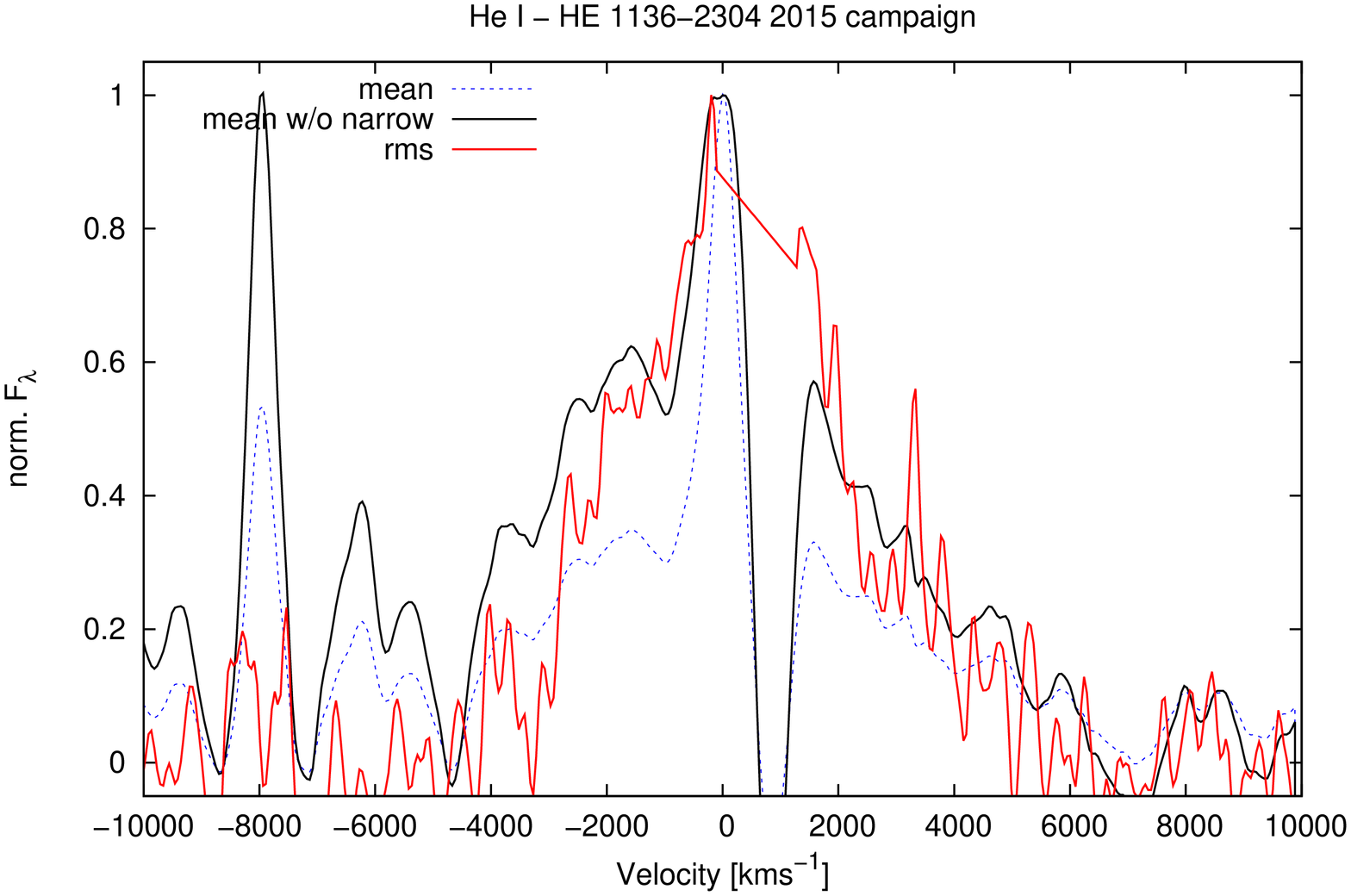}
  \caption{Normalized mean (blue), mean without a narrow component (black),
  and rms (red)
 line profiles of \ion{He}{i}\,$\lambda 5876$ in velocity space.}
  \label{velo_meanrms_he1.ps}
\end{figure}

\begin{figure}
  \includegraphics[height=9.cm,angle=-90]{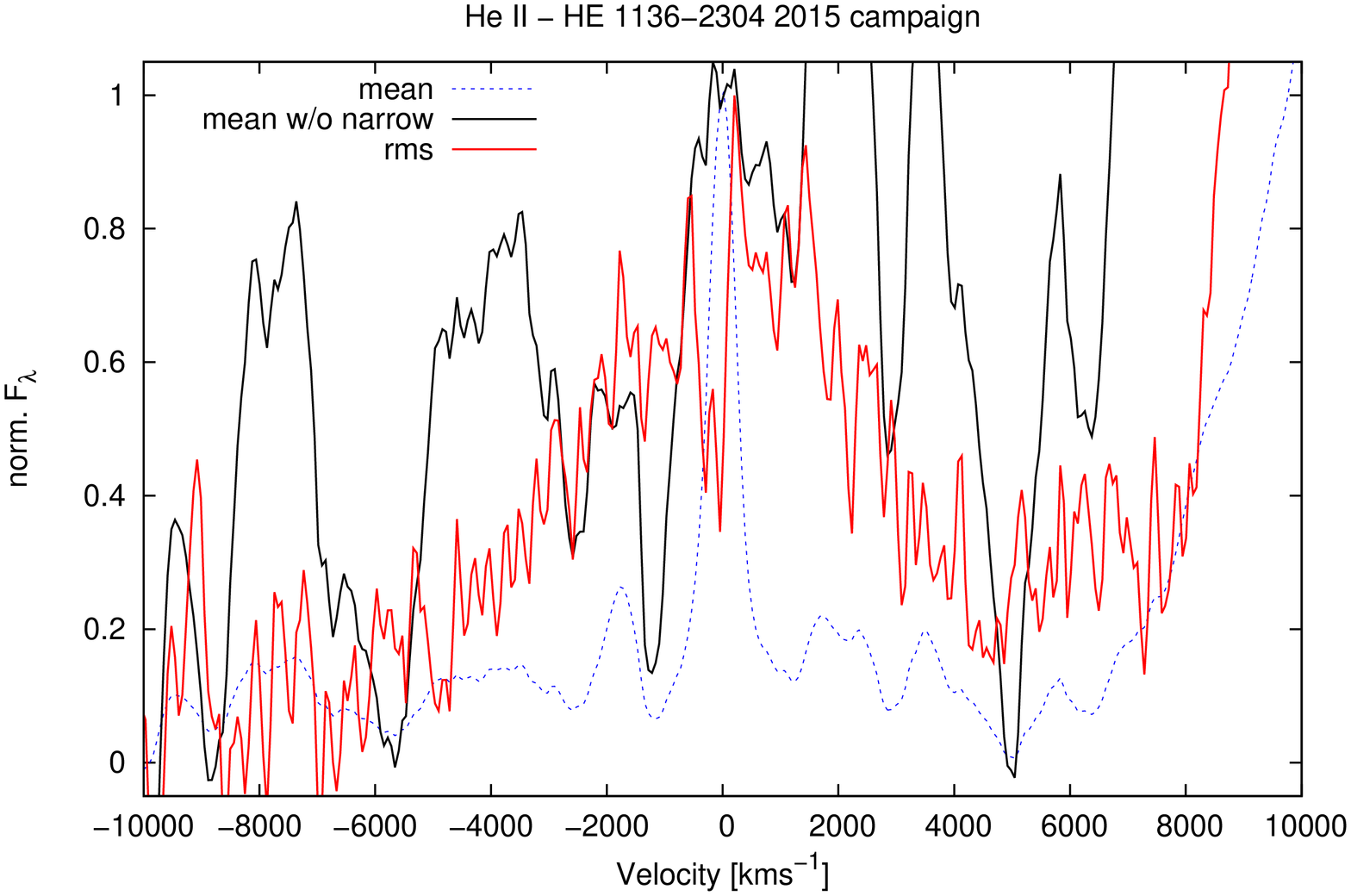}
  \caption{Normalized mean (blue), mean without a narrow component (black),
  and rms (red)
 line profiles of  \ion{He}{ii}\,$\lambda 4686$ in velocity space.}
  \label{velo_meanrms_he2.ps}
\end{figure}

\begin{figure}
 \includegraphics[height=9.cm,angle=-90]{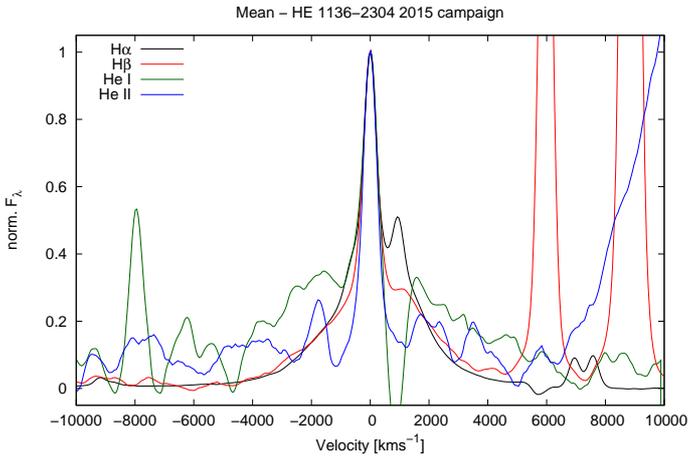}
   \caption{Normalized mean line profiles of
    H$\alpha$, H$\beta$, \ion{He}{i}\,$\lambda 5876$,
    and \ion{He}{ii}\,$\lambda 4686$.}
   \label{velo_mean_all.ps}
\end{figure}

\begin{figure}
 \includegraphics[height=9.cm,angle=-90]{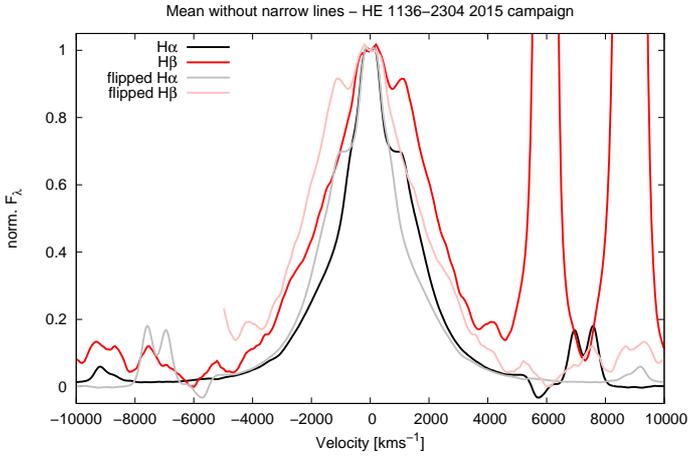}
  \caption{Normalized mean line profiles of
    H$\alpha$, H$\beta$ without a narrow component. In addition, we flipped
 the profiles at $v=0$ \kms.}
  \label{velo_mean_hahb_wo_narrow.ps}
\end{figure}

\begin{figure}
 \includegraphics[height=9.cm,angle=-90]{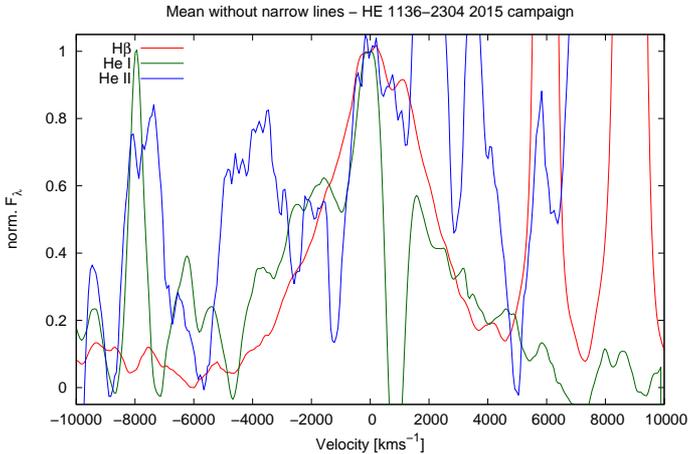}
  \caption{Normalized mean line profiles of
    H$\beta$,  \ion{He}{i}\,$\lambda 5876$,
    and \ion{He}{ii}\,$\lambda 4686$ without a narrow component.}
  \label{velo_mean_he_wo_narrow.ps}
\end{figure}

\begin{figure}
\vspace{-3mm}
 \includegraphics[height=9.cm,angle=-90]{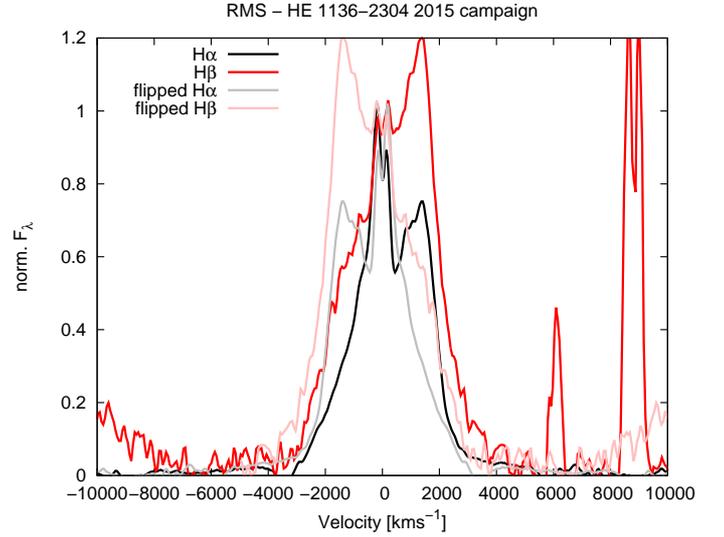}
  \caption{Normalized rms line profiles of
    H$\alpha$, H$\beta$. In addition, we flipped the profiles at
 $v=0$ \kms.}
  \label{velo_rms_hahb.ps}
\end{figure}

\begin{figure}
\vspace{-3mm}
 \includegraphics[height=9.cm,angle=-90]{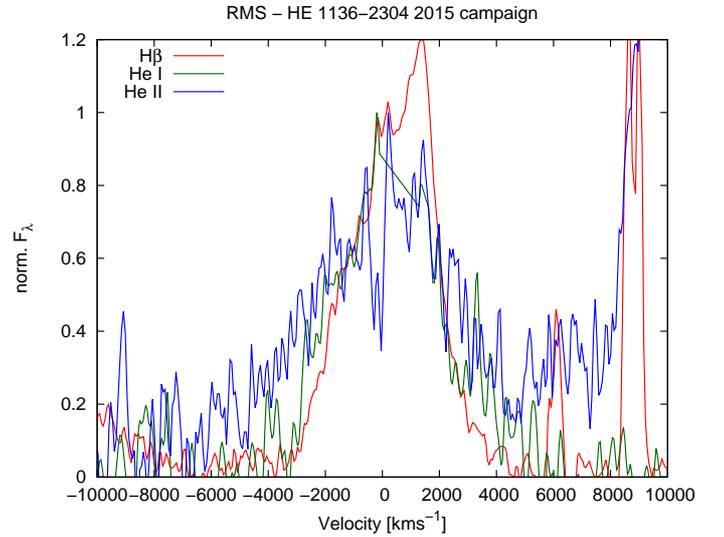}
  \caption{Normalized rms line profiles of
    H$\beta$, \ion{He}{i}\,$\lambda 5876$,
    and \ion{He}{ii}\,$\lambda 4686$.}
  \label{velo_rms_he.ps}
\end{figure}

We determined normalized mean and rms profiles of the Balmer
 and Helium lines after subtracting the 
continuum fluxes in each individual spectrum using the
continuum windows listed in Tab.~\ref{cont_boundaries}. 
Figs.~\ref{velo_meanrms_ha.ps} to \ref{velo_rms_he.ps}
show the mean and rms profiles of the Balmer lines
H$\alpha$ and H$\beta$ as well as those of the
He lines \ion{He}{i}\,$\lambda 5876$ and
\ion{He}{ii}\,$\lambda 4686$ in velocity space. 
The rms profiles illustrate the line profile variations during our campaign.
The constant narrow components disappear
almost completely in these rms profiles.     
However, the mean profiles contain
strong narrow line components in addition to their broad line components. We
subtracted the narrow Balmer and Helium line components in all the mean profiles
by subtracting a  scaled [\ion{O}{iii}]\,$\lambda$5007 line profile
as a template.
Furthermore, we subtracted the narrow
[\ion{N}{ii}]\,$\lambda 6584$ line in the mean H$\alpha$ profile and
removed the [\ion{Fe}{iii}]\,$\lambda 4861$ line in the
mean \ion{He}{ii}\,$\lambda 4686$ profile as well.
After subtracting these narrow components 
we compared the mean and rms profiles of the broad
emission lines with each other in a more accurate way. 

We present the normalized mean profiles
(with and without narrow components) and rms profiles
for all four Balmer and Helium emission lines in
 Figs.~\ref{velo_meanrms_ha.ps} to
\ref{velo_meanrms_he2.ps}.
%
%
%
A comparison of all normalized
mean profiles including their narrow components is shown
in Fig.\ref{velo_mean_all.ps}.
We then present a comparison of the mean profiles after subtracting
the narrow components; this comparison is shown in
Fig.\ref{velo_mean_hahb_wo_narrow.ps} for the mean
profiles of H$\alpha$ and H$\beta$. For getting a hint on line asymmetries
their flipped profiles at $v=0$ \kms are depicted as well.
Fig.\ref{velo_mean_he_wo_narrow.ps} presents the mean Helium
profiles after subtracting their narrow components. The profile
of H$\beta$ has also been added for comparison purposes.
Fig.\ref{velo_rms_hahb.ps} shows a comparison of the
H$\alpha$ and H$\beta$ rms line profiles normalized
with respect to their central component.
Again their flipped profiles at $v=0$ \kms are depicted
to highlight the additional strong red components
and the weak blue component.
Finally, Fig.\ref{velo_rms_he.ps} gives the rms profiles of the Helium
lines. Again, the H$\beta$ line profile
has been added for comparison.

We determined the linewidths (FWHM) of the mean
and rms line profiles of all the Balmer and Helium lines.
These measurements were performed with respect to the 
maximum of the central component and are listed 
 in Tab.~\ref{line_widths}.
In addition, we parameterized the linewidths of the rms profiles by
their line dispersion $\sigma_\text{line}$ (rms widths) 
(Fromerth \& Melia\citealt{fromerth00}; Peterson et al.\citealt{peterson04}).
\begin{table}
\centering
\tabcolsep+2mm
\caption{Balmer and Helium linewidths: FWHM of the mean
and rms line profiles as well as the line dispersion $\sigma_\text{line}$ 
of the rms profiles.}
\begin{tabular}{lccc}
\hline 
\noalign{\smallskip}
Line & FWHM (mean) & FWHM (rms) & $\sigma_{line}$ (rms) \\
     & [\kms] & [\kms] & [\kms] \\
(1)  & (2) & (3) & (4) \\ 
\noalign{\smallskip}
\hline
\noalign{\smallskip}
H$\alpha$                        & 2523$\pm$  100 & 2668 $\pm$ 150 &  1816 $\pm$ 150 \\
H$\beta$                         & 4063$\pm$  100 & 3791 $\pm$ 150 &  1767 $\pm$ 150 \\
\ion{He}{i}\,$\lambda 5876$      & 4557$\pm$  400 & 4131 $\pm$ 400 &  2098 $\pm$ 400 \\
\ion{He}{ii}\,$\lambda 4686$     & 5070$\pm$ 1000 & 5328 $\pm$ 500 &  2962 $\pm$ 500 \\
\noalign{\smallskip}
\hline 
\end{tabular}
\label{line_widths}
\end{table}
The Balmer line profiles (mean and rms) exhibit linewidths (FWHM) between
2500 and 4000 \kms. H$\alpha$ shows the narrowest broad line profiles.
The Helium lines (\ion{He}{i}\,$\lambda 5876$  and \ion{He}{ii}\,$\lambda 4686$)
are always broader than the Balmer lines by 500 to 1500 \kms{}.
Furthermore, the outer blue wing is more extended in the higher ionized He lines
in comparison to the Balmer lines
(Figs.\ref{velo_rms_hahb.ps},  \ref{velo_rms_he.ps}).
This might be an indication of an additional outflowing
 component in the inner broad-line region (BLR).

An additional strong red component emerges in the
rms and mean profiles of the H$\alpha$ and H$\beta$ lines
at 1380 \kms 
and varies with a larger amplitude than the rest of the line profile.
Furthermore, this component shows a relatively stronger
variation in H$\beta$ 
in comparison with H$\alpha$
(see Fig.\ref{velo_rms_hahb.ps}) by a factor of 1.5.
An indication of the existence of this red component is present in the 
 \ion{He}{ii}\,$\lambda 4686$ line
as well (Fig.\ref{velo_rms_he.ps}). No clear indication of this component
is visible in the red wing of the \ion{He}{i}\,$\lambda 5876$ line as
that region coincides with the NaD absorption.
An additional blue component, nearly symmetrical to the red component, appears in the rms profile of H$\beta$ at around -1400~\kms.
(Fig.\ref{velo_rms_hahb.ps}). However, this blue
component is by far weaker than the red component.  
The existence of blue and red components in the line profiles --
in addition to the central component -- are an indication
that the broad line-emitting region is connected with an accretion disk
structure.

\subsection{CCF analysis of the integrated broad emission lines}

\subsubsection{Based on SALT spectra}

\begin{figure*}
\vspace{-3mm}
\centering
 \includegraphics[width=10.cm,angle=-90]{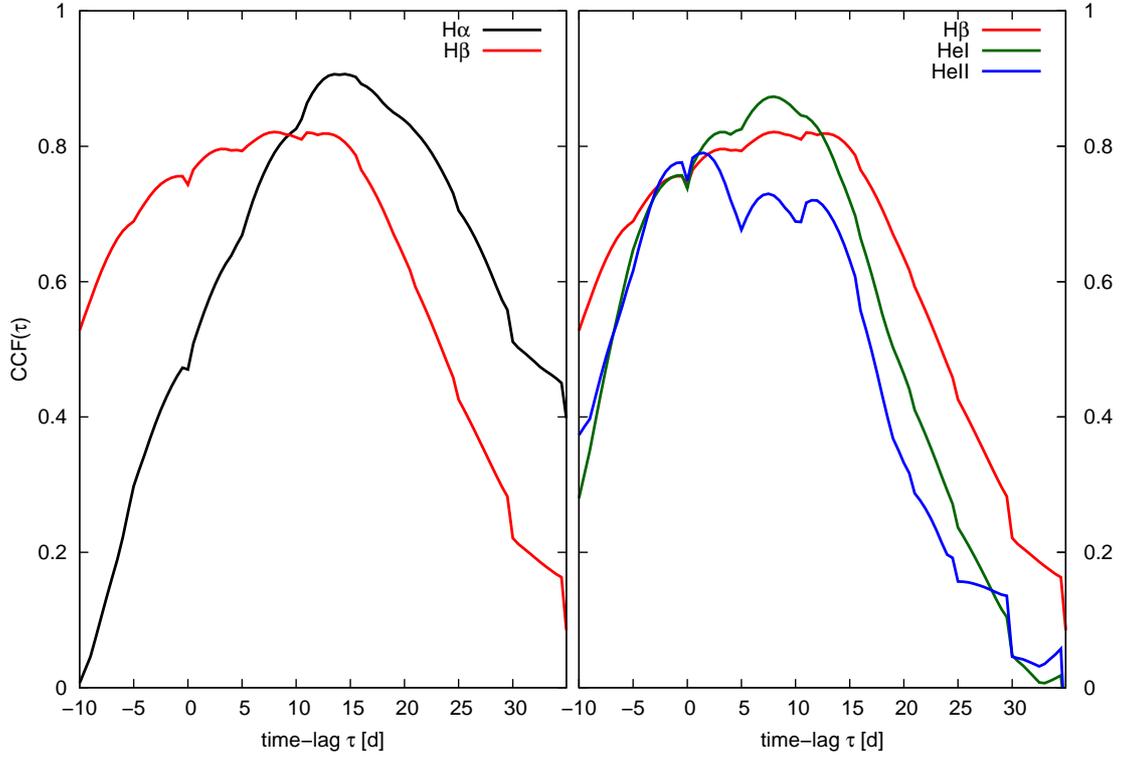}
  \caption{Cross-correlation functions of the integrated
    H$\alpha$, H$\beta$, \ion{He}{i}\,$\lambda 5876$,
    and \ion{He}{ii}\,$\lambda 4686$ lines with respect to the continuum
   at  4570\,\AA{}.}
  \label{CCF_SALT_ochm.ps}
\end{figure*}

The mean distances of the broad emitting line regions to the central
ionizing source can be determined by correlating
the broad emission line light curves
with the light curves of the ionizing continuum flux.
Normally, an optical continuum light curve is
used as a surrogate for the ionizing flux light curve. 
An interpolation cross-correlation function
method (ICCF) has been developed by Gaskell \& Peterson\cite{gaskell87}
to calculate the delay of the individual line light curves 
with respect to the continuum light curve.
We generated our own ICCF code 
(Dietrich \& Kollatschny\citealt{dietrich95}) based on similar assumptions
in the past.
For this study, we correlated the light curves
of the integrated Balmer (H$\alpha$, H$\beta$)   
 and Helium lines (\ion{He}{i}\,$\lambda 5876$, \ion{He}{ii}\,$\lambda 4686$)
 with the continuum light curve at 4570\,\AA{} using our method. The
derived ICCF($\tau$)
are presented in Fig.\ref{CCF_SALT_ochm.ps}.
\begin{figure*}
\vspace{-3mm}
\centering
 \includegraphics[width=10.cm,angle=-90]{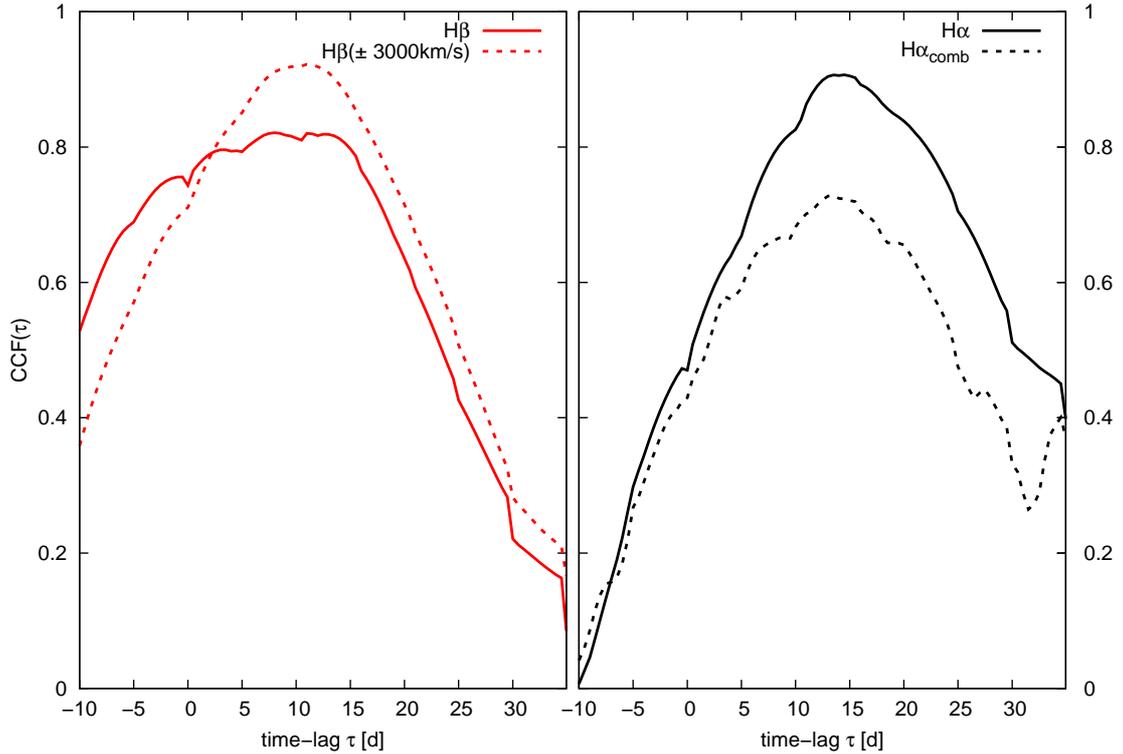}
  \caption{Cross-correlation functions of the integrated H$\beta$ line and the
    inner part only at $\leq \pm$ 3000 \kms{} (left). 
    CCFs based on the integrated H$\alpha$ line based on SALT spectra and on additional data from the narrowband photometry
    (NB$_{670}$ filter; see Paper 1) with respect to the continuum
     at 4570\,\AA{} \textbf(right).
}
  \label{CCF_3000_ochm.ps}
\end{figure*}
We determined the centroids of these ICCF, $\tau_\text{cent}$
by using only those parts of the CCFs above 80\% of the peak value
{\bf $r_\text{max}$}.
A threshold value of 0.8 $r_\text{max}$ is generally a good choice
as had been shown by, for example, Peterson et al.\citealt{peterson04}. 
We derived the uncertainties in our cross-correlation results by
calculating the cross-correlation lags a large number of times using
a model-independent Monte Carlo method known as
flux redistribution/random subset selection (FR/RSS).
This method has been described
by Peterson et al.\cite{peterson98}.
Here the error intervals correspond to 68\% confidence levels. 
The final results of the ICCF analysis are given in Tab.~\ref{CCF_1D}.
\begin{table}
\centering
\tabcolsep+16mm
\caption{Cross-correlation lags of the Balmer and Helium line light curves
     with respect to the 4570\,\AA\ continuum light curve.}
\begin{tabular}{lr}
\hline 
\noalign{\smallskip}
Line & \multicolumn{1}{c}{$\tau$} \\
     & \multicolumn{1}{c}{[days]}\\
\noalign{\smallskip}
\hline
\noalign{\smallskip}
\Ha                          &   $15.0^{+4.2}_{-3.8}$\\[.7ex]
\Hb                          &   $7.5^{+4.6}_{-5.7}$\\[.7ex]
\ion{He}{i}\,$\lambda 5876$  &   $7.3^{+2.8}_{-4.4}$\\[.7ex]
\ion{He}{ii}\,$\lambda 4686$ &   $3.0^{+5.3}_{-3.7}$\\[.7ex]
\noalign{\smallskip}
\hline 
\end{tabular}
\label{CCF_1D}
\end{table}
The delay of
the line-averaged BLR size 
of the integrated Balmer lines
with respect to the continuum light
curve at 4570\,\AA\ 
corresponds to $15.0^{+4.2}_{-3.8}$ and
$7.5^{+4.6}_{-5.7}$ light days in the H$\alpha$ and H$\beta$ lines,
respectively.  
The \ion{He}{i}\,$\lambda 5876$ shows a delay of $7.3^{+2.8}_{-4.4}$
days while the delay of
the integrated \ion{He}{ii}\,$\lambda 4686$ line   
corresponds to $3.0^{+5.3}_{-3.7}$ light days.

The CCF of H$\beta$ is very broad in comparison
to H$\alpha$ (see Fig.\ref{CCF_SALT_ochm.ps}).
We discuss in section 3.5 how individual segments
of the emission lines exhibit different delays with respect to the 
ionizing continuum.
The line wings originate closer to the
central ionizing zone while the central line regions originate
at larger distances from the central ionizing zone.
We carried out additional tests whether the integrated inner
line profile segments 
(at $\pm$3000\,\kms) give similar or rather slightly larger CCF lags
in comparison to the total line profiles. 
For H$\beta$ (at $\pm$3000\,\kms{})  we got a delay of
ten light days (see Fig.\ref{CCF_3000_ochm.ps})
in comparison to the delay of 7.5 light days for 
the integrated profile.

\subsubsection{Based  on photometric light curves in combination with 
SALT spectra}

We published the continuum data of HE\,1136-2304
taken by {\it Swift} in the B band and 
additional B-band observations taken with the
MONET North and South telescopes and with the Bochum
telescopes at Cerro Armazones in Zetzl et al.\cite{zetzl18}. 
Subsequently, we generated a combined B-band light curve
B$_\text{Bochum}$ in combination with our SALT spectral data.
Furthermore, we created a combined H$\alpha$ light curve
H$\alpha_\text{comb}$ based 
on the SALT spectra and the narrowband NB$_{670}$ 
light curve taken with the Bochum telescope.
In a next step we carried out some tests to ascertain whether CCFs 
based on various combinations of the continuum light curves
and based on combined H$\alpha$ light curves
produce similar results.

\begin{figure}
\centering
\includegraphics[height=9.cm,angle=-90.0]{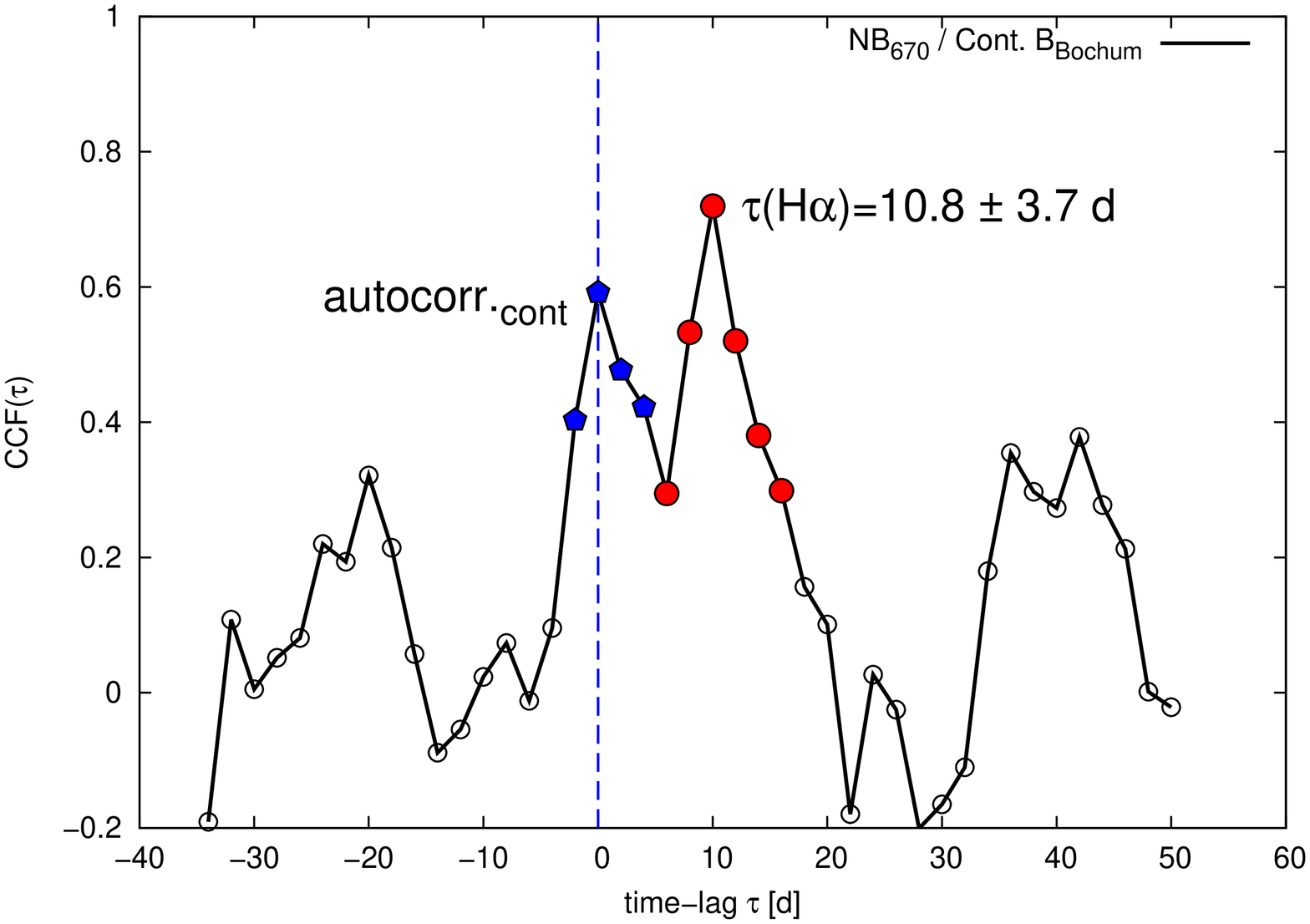}
\caption{Discrete correlation function of the H$\alpha$ line based on
 the Bochum narrowband filter
flux (NB$_{670}$) with respect to the Bochum B-band flux.} 
\label{DCF.ps}
\end{figure}

In Tab.~\ref{CCF_Ha_cont} we present \textbf{1)}
the cross-correlation lags of the spectroscopically obtained H$\alpha$
light curve correlated
with the combined B-band light curve and  \textbf{2)}
the combined H$\alpha_\text{comb}$ light curve correlated with the
4570\,\AA\ continuum light curve based on the SALT spectra.
We also derived \textbf{3)} the H$\alpha$ lag from the Bochum light curves 
in the B and NB$_{670}$ bands; the NB$_{670}$ band entirely covers the broad
H$\alpha$ line.
Because the narrowband filter contains the contribution from the AGN
continuum underneath H$\alpha$  in addition to H$\alpha$, 
we might expect two peaks in the cross correlation (see also 
Chelouche \& Daniel \citealt{chelouche12}).
To calculate the cross correlation we used   
the discrete correlation function (DCF; Edelson \& Krolik \citealt{edelson88}) 
with steps of $\delta \tau = 2$~days.
In comparison to the spectroscopic SALT light curves, the number of data points
in this case is sufficiently high to allow for such a DCF analysis, which
in general yields better results in case of unevenly sampled data.
This cross correlation indeed shows two clearly distinct peaks 
separated by a deep minimum in between
(Fig. \ref{DCF.ps}). 
The peak at lag $\tau \approx 0$~days (denoted with blue dots) comes from 
the cross correlation of the AGN continuum in the B band with the continuum 
underneath the H$\alpha$ line; it is virtually an autocorrelation.
The second even brighter peak at $\tau \approx 11$~days  
(denoted with red dots) 
comes from the lag between the B-band continuum and H$\alpha$.
The derived  H$\alpha$ lag (using the red data points) agrees within 
the errors with the H$\alpha$ lags derived from the SALT spectra.
Regarding the error limits, the derived H$\alpha$ delays 
(11 up to 17 light days) are 
in good agreement with the delay of 15 light days based on the SALT
spectra alone.
\begin{table}
\centering
\tabcolsep+8.5mm
\caption{Cross-correlation lags of the H$\alpha$ light curve with the combined
B-band light curve, the combined H$\alpha$ light curve with the
continuum light curve based on the SALT data,
and the H$\alpha$ light curve based on the
Bochum NB$_{670}$ date with the continuum light curve based on the Bochum B-band data.}
\begin{tabular}{lr}
\hline 
\noalign{\smallskip}
Line & \multicolumn{1}{c}{$\tau$} \\
     & \multicolumn{1}{c}{[days]}\\
\noalign{\smallskip}
\hline
\noalign{\smallskip}
\Ha$_\text{SALT}$\,\,/\,\,Cont. B$_\text{comb}$       &   $16.7^{+2.3}_{-5.5}$\\[.7ex]
\Ha$_\text{comb}$\,\,\,/\,\,Cont. 4570\,\AA\         &   $13.7^{+2.9}_{-5.0}$\\[.7ex]
NB$_{670}$\,\,\,\,\ /\,\,Cont. B$_\text{Bochum}$      &   $10.8^{+3.7}_{-3.7}$\\[.7ex]
\noalign{\smallskip}
\hline 
\end{tabular}
\label{CCF_Ha_cont}
\end{table}

\begin{figure}
\centering
\includegraphics[height=9.cm,angle=-90]{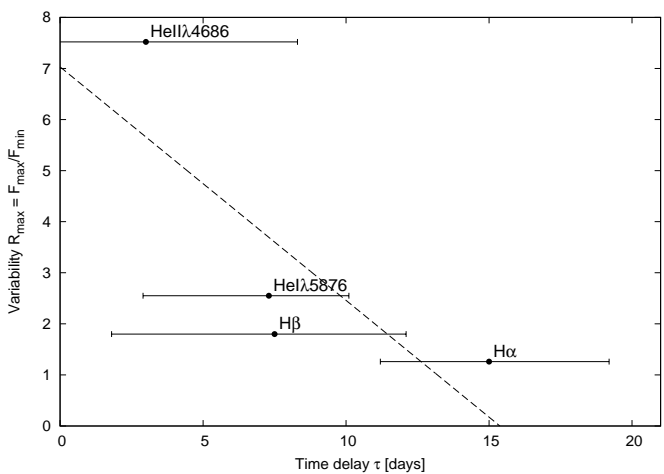}
\caption{Variability amplitude of the integrated emission lines
 as a function of their time delay $\tau$ (i.e., distance to
 the center). The dashed line indicates the linear fit to the data.}
\label{rmax_vs_tau.ps}
\end{figure}

\begin{figure}
\centering
\includegraphics[height=9.cm,angle=-90]{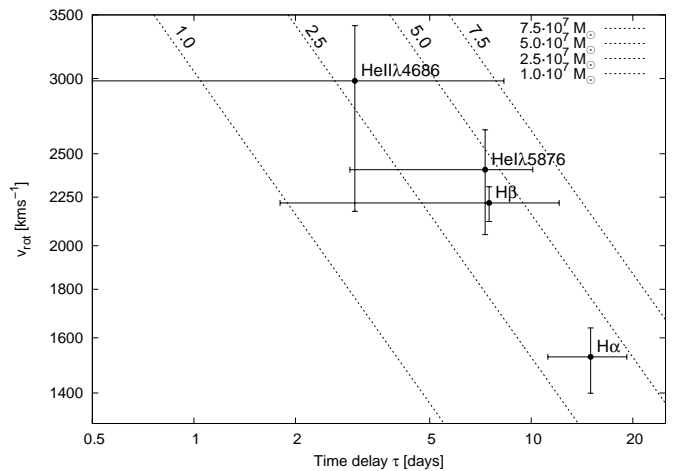}
\caption{Linewidth of the emission lines (FWHM $v_{\text{rot}}$) as a function of their
 time delay $\tau$ (i.e., distance to
 the center). The dashed lines correspond to virial masses of
1.0, 2.5, 5.0, and 7.5$\times 10^7$ M$_{\odot}$.}
\label{vrot_vs_tau.ps}
\end{figure}

It has been demonstrated for other AGN (e.g., 3C120; 
Kollatschny et al.,\citealt{kollatschny14})
that the variability amplitude of the
integrated emission lines is inversely correlated with the distance
of the line-emitting regions to the central ionizing source.
Fig.\ref{rmax_vs_tau.ps} shows that this relation is valid for
the broad lines in HE\,1136-2304 as well. 
Furthermore, it has been shown that there is a radial stratification 
with respect to the BLR linewidths (FWHM;
e.g., Kollatschny\citealt{kollatschny03}). 
The higher ionized lines show broader linewidths (FWHM) and originate
closer to the center as shown in Fig.\ref{vrot_vs_tau.ps}. 
For this figure we used the corrected
rotational velocities $v_\text{rot}$ as presented in
Tab.~\ref{acc_str_h/r.ps} (see section 3.6).
For comparison with our measurements, in Fig. 19 we show 
the expected relations between distance and
 linewidth for multiple black hole masses
based on the mass formula given in section 3.4.

\subsection{Central black hole mass}

The masses of the central black holes in AGN can be
estimated from
the width of the broad emission line profiles, based on the assumption 
that the gas dynamics are dominated by the
central massive object, by evaluating
$M = f\,c\,\tau_\text{cent}\,\Delta\,v^{2}\, G^{-1}  $.
It is necessary to know the distance of the line-emitting region.
Characteristic distances of the individual line-emitting regions
are given by the centroid $\tau_\text{cent}$
 of the individual cross-correlation
functions of the emission-line variations relative to the continuum variations
 (e.g., Koratkar \& Gaskell\citealt{koratkar91};
Kollatschny \& Dietrich\citealt{kollatschny97}).
The characteristic velocity $\Delta v$ of the emission-line
regions can be estimated from the FWHM of the rms profiles
or from the line dispersions $\sigma_\text{line}$.

The scaling factor $f$ in the equation above is on the order
of unity and depends
on the kinematics, structure, and orientation of the BLR.
This scaling factor may differ from galaxy to galaxy,
for example, depending on whether we see the
central accretion disk including the BLR from the edge or face-on. 
We compared the central black hole mass value of HE\,1136-2304 
with values of the black hole masses derived for other AGN
and adopt a mean value of $f=5.5$
(e.g., Onken et al.\citealt{onken04}; 
Grier et al.\citealt{grier12}). 
This f-value might be too high by a factor of two when
comparing the black hole masses
with inactive galaxies (Graham et al.\citealt{graham11}).
Nevertheless, using $f=5.5$, 
we calculate a mean black hole mass (see Tab.~\ref{bh_masses}, col.~3) of 
\[ M = 3.5 \pm 2.8 \times 10^{7} M_{\odot}, \]
based on the derived delays of the integrated Balmer and Helium lines
(see Tab.~\ref{CCF_1D}) and on the line dispersions $\sigma_\text{line}$
(see Tab.~\ref{line_widths}).
All BH masses based on the individual lines agree with each other within
the error limits.

\begin{table}
\centering
\tabcolsep+2.8mm
\caption{Black hole masses based on $v_{\text{rot}}$ , $\sigma_\text{line}$
 (rms width), and FWHM (rms width).}
\begin{tabular}{lccc}
\hline 
\noalign{\smallskip}
Line & M$_{\text{BH},V_{\text{rot}}}$ &  M$_{\text{BH},\sigma_\text{line}}$ &M$_{\text{BH, FWHM}}$ \\
     &  & [$10^7 M_{\odot}$] &  \\
(1)  & (2) & (3) & (4) \\ 
\noalign{\smallskip}
\hline
\noalign{\smallskip}
H$\alpha$                    &3.8 $\pm{}$ 1.2 & 5.3 $\pm{}$ 1.7 & 11.5 $\pm{}$ 3.5\\
H$\beta$                     &4.0 $\pm{}$ 3.0 & 2.5 $\pm{}$ 2.0 & 11.6 $\pm{}$ 8.8\\
\ion{He}{i}\,$\lambda 5876$  &4.5 $\pm{}$ 3.0 & 3.5 $\pm{}$ 2.5 & 13.4 $\pm{}$ 8.5\\
\ion{He}{ii}\,$\lambda 4686$ &2.9 $\pm{}$ 5.3 & 2.8 $\pm{}$ 5.1 & 9.1 $\pm{}$ 16.2\\
\noalign{\smallskip}
\hline
\noalign{\smallskip}
mean                           &3.8 $\pm{}$ 3.1 & 3.5 $\pm{}$ 2.8 & 11.4 $\pm{}$ 9.2  \\ 
\hline 
\end{tabular}
\label{bh_masses}
\end{table}

Using the linewidths FWHM (rms) (see Tab.~\ref{line_widths}) we
calculated a mean black hole mass of
$M = 11.4 \pm 9.3 \times 10^{7} M_{\odot}$
(Tab.~\ref{bh_masses}, col.~4).
However, in that case we did not correct for the contribution of
turbulent motions to the width of the line profiles; this is covered in 
section 3.7.
After correcting the emission linewidths (FWHM) for their
contribution of turbulent motions 
(Tab.~\ref{acc_str_h/r.ps}, col.~4),
 we derive a mean black hole mass of
\[ M = 3.8 \pm 3.1 \times 10^{7} M_{\odot} \]
(Tab.~\ref{bh_masses}, col.~2).
Again, all individual BH masses based on this method agree with each other
within the error limits.

\subsection {Two-dimensional CCFs of  Balmer (H$\alpha$, H$\beta$)
 and Helium\,I, II line profiles}

We first calculated the cross-correlation lags of the integrated 
Balmer and helium lines with respect to the continuum as mentioned
in section 3.3. 
Now we investigate the profile variations of
these lines in more detail by calculating the lags of individual line segments.
The way we proceed has been described before in our studies of
line profile variations in
Mrk\,110 (Kollatschny \& Bischoff\citealt{kollatschny02}; 
Kollatschny\citealt{kollatschny03}), Mrk\,926
(Kollatschny \& Zetzl\citealt{kollatschny10}), and 3C120   
(Kollatschny et al.\citealt{kollatschny14}).

\begin{figure}
\includegraphics[width=65mm,angle=-90]{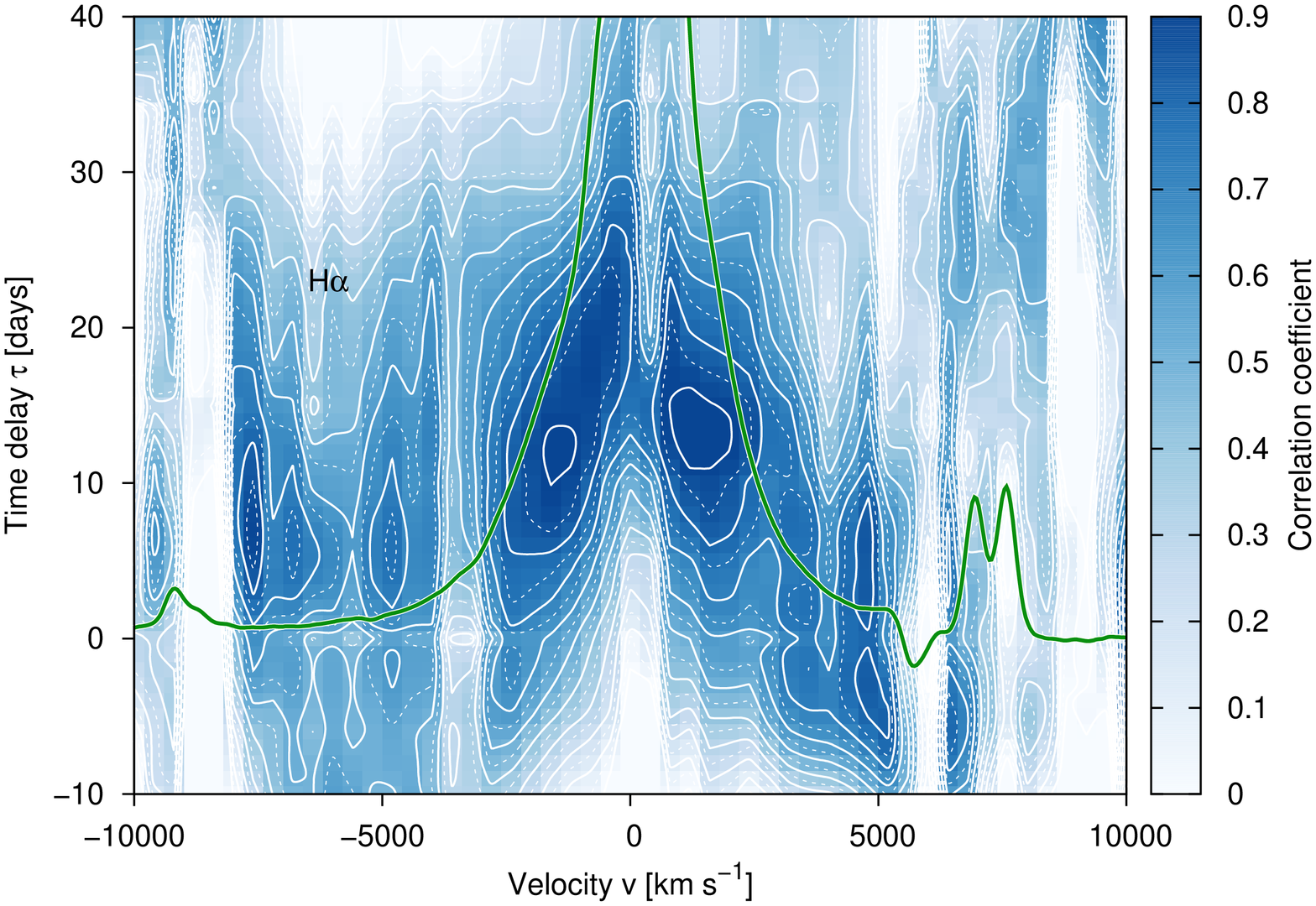}
  \caption{Two-dimensional CCF($\tau$,$v$) showing the correlation coefficient
 of the H$\alpha$ line  
segment light curves with the continuum light curve
as a function of velocity and time delay (blue scale).
Contours of the correlation coefficients are plotted at levels
0.0 to 0.9 every 0.05 (white lines).
The green line shows the line profile of the mean spectrum.
}
   \label{ccf2d_ha.ps}

\includegraphics[width=65mm,angle=-90]{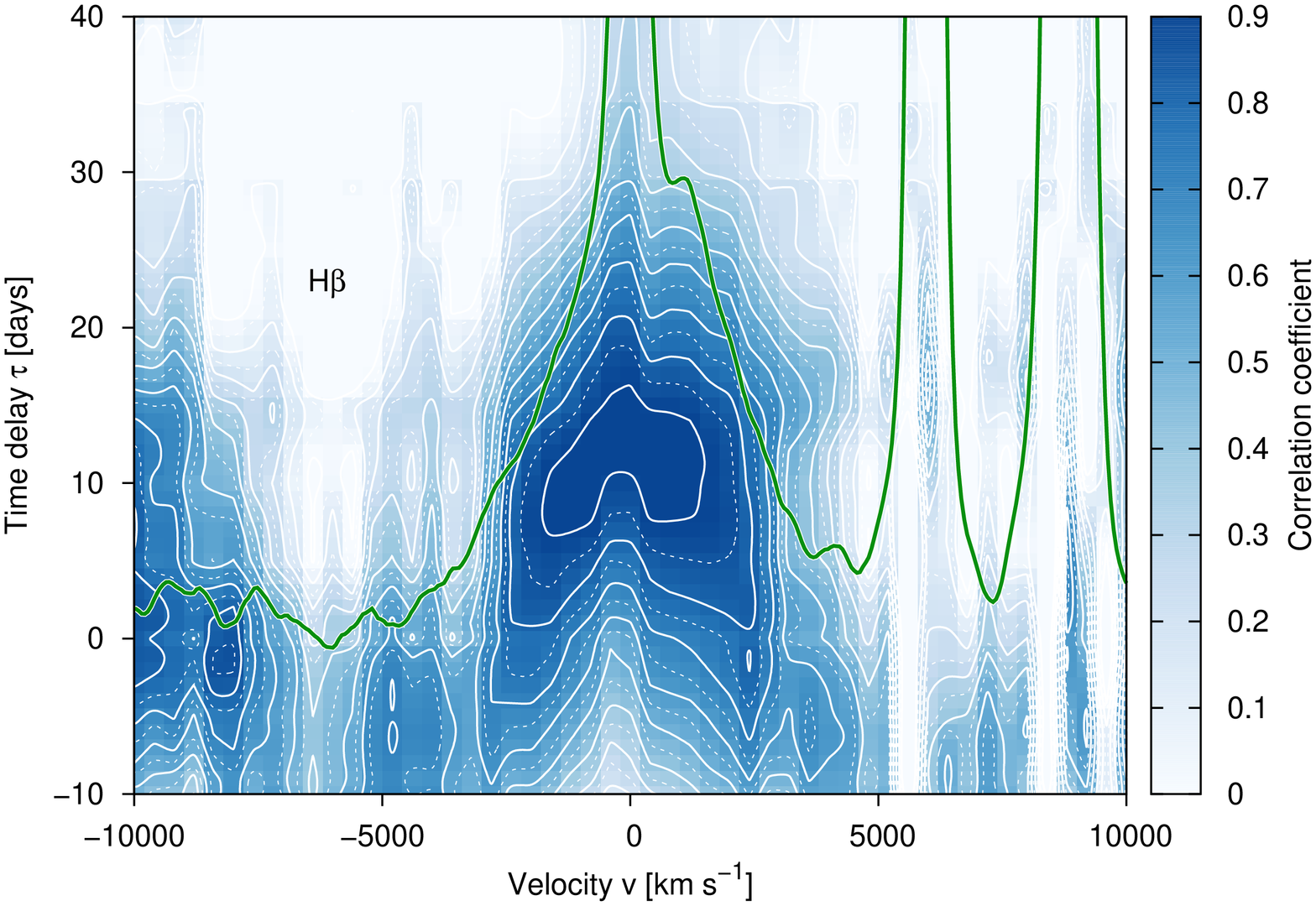}
  \caption{Two-dimensional CCF($\tau$,$v$) showing the correlation coefficient
 of the H$\beta$ line  
segment light curves with the continuum light curve
as a function of velocity and time delay (blue scale).
Contours of the correlation coefficients are plotted at levels
0.0 to 0.9 every 0.05 (white lines).
The green line shows the line profile of the mean spectrum.
}
   \label{ccf2d_hb.ps}
\end{figure}

\begin{figure}
\includegraphics[width=65mm,angle=-90]{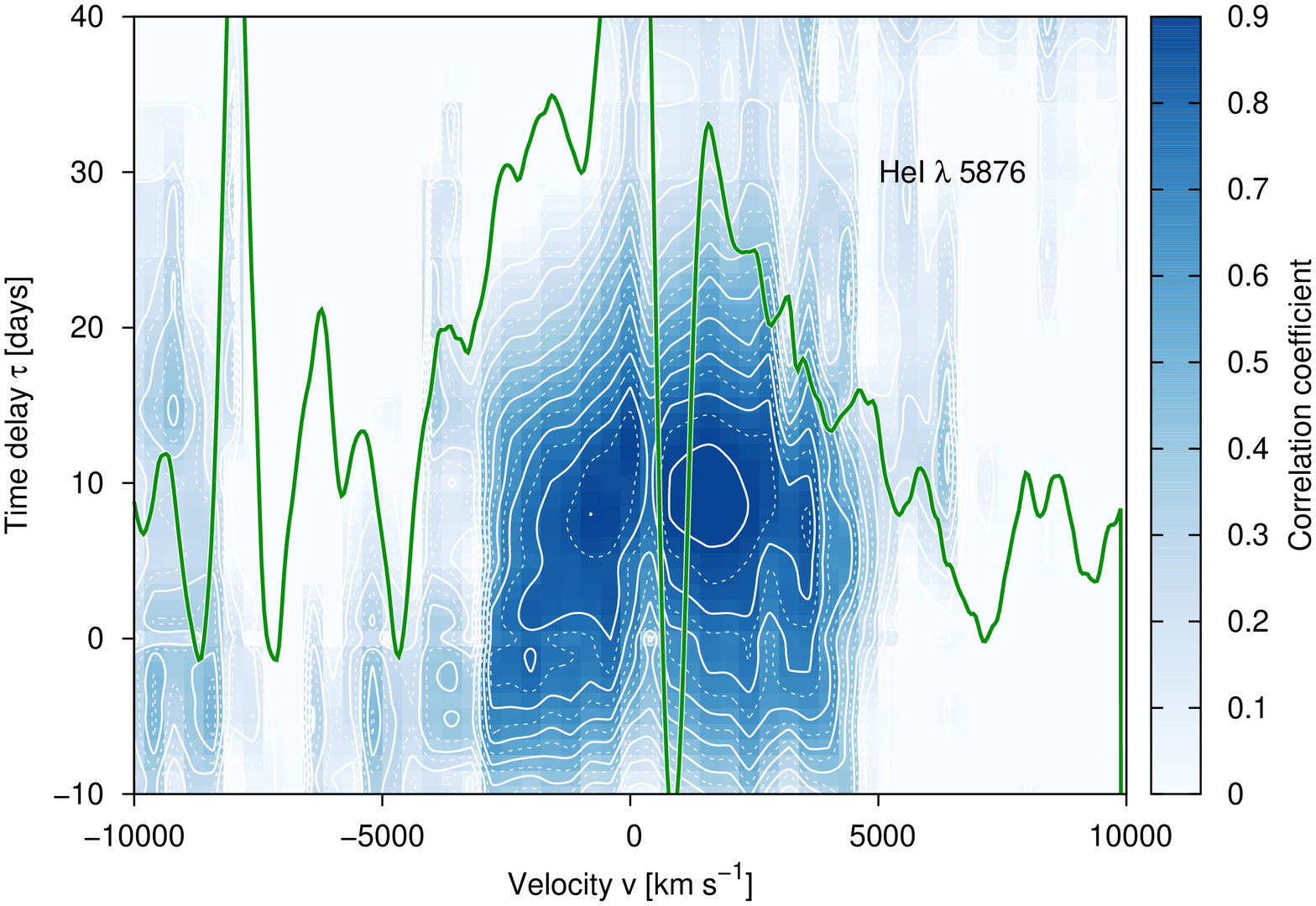}
  \caption{Two-dimensional CCF($\tau$,$v$) showing the correlation coefficient
 of the \ion{He}{i}\,$\lambda 5876$ line  
segment light curves with the continuum light curve
as functions of velocity and time delay (blue scale).
Contours of the correlation coefficients are plotted at levels
0.0 to 0.9 every 0.05 (white lines).
The green line shows the line profile of the mean spectrum.
}
   \label{ccf2d_he1.ps}
\includegraphics[width=65mm,angle=-90]{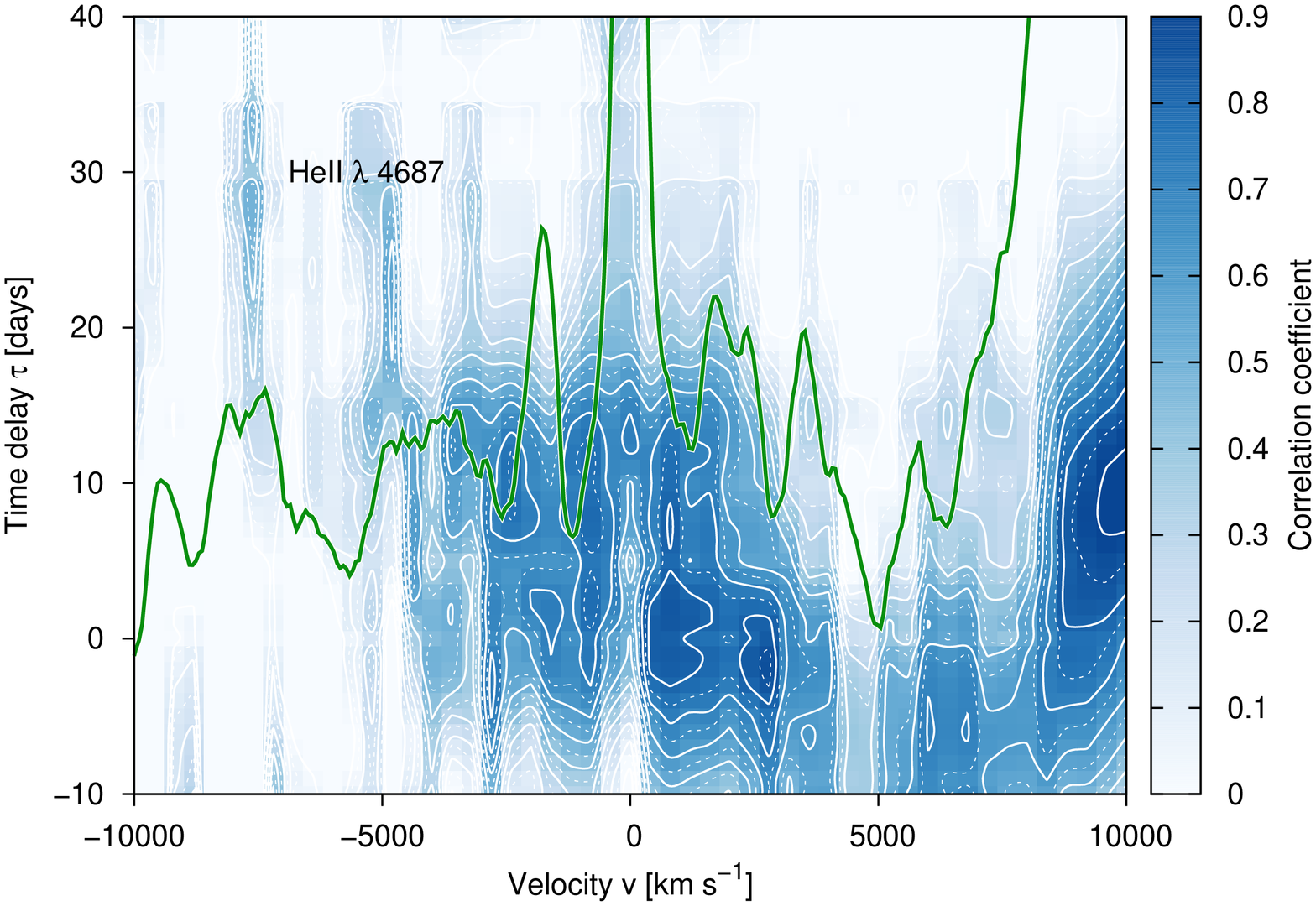}
  \caption{Two-dimensional CCF($\tau$,$v$) showing the correlation coefficient
 of the \ion{He}{ii}\,$\lambda 4686$ line  
segment light curves with the continuum light curve
as functions of velocity and time delay (blue scale).
Contours of the correlation coefficients are plotted at levels
0.0 to 0.9 every 0.05 (white lines).
The green line shows the line profile of the mean spectrum.
}
   \label{ccf2d_he2.ps}
\end{figure}

We sliced the velocity profiles of the Balmer and Helium lines  
into velocity segments with a width of $\Delta$$v = 400$ \kms{}, which 
 corresponds to the spectral resolution of our spectra.
A central line segment was integrated in the
velocity range -200 \kms $\le v \le$ 200 \kms.
Afterward, we measured the intensities of all subsequent
velocity segments 
from $v = -9\,800$ to $+9\,800$\,\kms 
and compiled their light curves. Light curves of the central Balmer
and Helium line segments and selected blue and red segments
at $800$, $2000$, and $4000$\,\kms{}
are shown in Figs.~\ref{lc_seg_ha_select.ps} to \ref{lc_seg_he2_select.ps}.
For comparison, the light curve of the continuum flux at 4570\,\AA\
is given in all these figures as well.
We computed the maximal correlation coefficient 
and time delay $\tau_\text{cent}$ of all
line segment ($\Delta{}v=400$\,\kms) light curves of the Balmer and
Helium lines
with respect to the 4570\,\AA{} continuum flux light curve.
The derived time delays of the segments are shown in 
Figs.~\ref{ccf2d_ha.ps} to \ref{ccf2d_he2.ps}
as functions of distance to the line center (blue scale).
The white lines in  Figs.~\ref{ccf2d_ha.ps} to \ref{ccf2d_he2.ps} delineate the
contour lines of the correlation coefficient
at different levels.
The green line shows the line profile of the mean spectrum for comparison.
%
%

The following statements can be made based on these figures. 
There is a general trend that the Helium line response is faster than the 
response of the Balmer lines, as already known from the integrated lines.
The velocity-delay maps are very symmetric with respect to the line center. 
The light curves of the emission line centers
are delayed by 10 to 20 days with respect
to the continuum variations, while the
outer line wings at distances of $+/-$2000 to $+/-$3000 \kms{} 
respond much faster to continuum variations 
and only show a delay of 0 to 10 days.
The delay in the outer line wings at distances of $+/-$4000 \kms{} 
is even negative with respect to the optical continuum.
It has been discussed in Zetzl et al.(\citealt{zetzl18}) that the observed optical continuum is delayed with respect to the ionizing continuum in
the UV and X-ray bands.
The outer blue wing of  the H$\beta$ line shortward of $-6000$\,\kms)
is blended with the red wing of the HeII$\lambda$4686 line (see Fig.2 as well).

\subsection{Vertical BLR structure in HE\,1136-2304}

Information about the BLR structure in Seyfert 1 galaxies
can be derived from the profiles of the broad emission lines
together with variability studies (Kollatschny
 \& Zetzl\citealt{kollatschny11,kollatschny13a,kollatschny13b,kollatschny13c}).
The broad emission line profiles can be
parameterized by the ratio of their full-width at half maximum (FWHM)
 to their line dispersion $\sigma_{\mathrm{line}}$.
We were able to show that there exists a general relation between
the FWHM and the linewidth ratio 
FWHM/$\sigma_{\mathrm{line}}$. 
The linewidth FWHM primarily reflects the line broadening of the
intrinsic Lorentzian
profiles due to rotational motions of the broad line gas.
The intrinsic Lorentzian
profiles themselves are associated with turbulent motion
(see also Goad et al.\citealt{goad12})
and different emission lines turn out to 
exhibit, on average, characteristic turbulent 
velocities within a narrow range
(Kollatschny \& Zetzl\citealt{kollatschny11}).

\begin{figure}[t]
\includegraphics[width=62mm, angle=270]{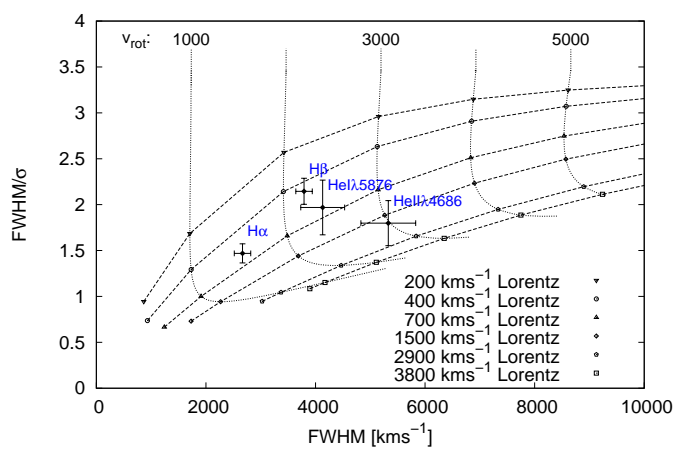}
  \caption{Observed and modeled linewidth ratios
 FWHM/$\sigma_{\mathrm{line}}$ vs. linewidth FWHM in  HE\,1136-2304.
 The dashed curves represent the corresponding theoretical
linewidth ratios based on rotational line-broadened
 Lorentzian profiles (FWHM = 200 to
3800\,\kms). The rotation velocities reach
from 1000 to 5000\,\kms
(curved dotted lines from left to right).}
   \label{fwhmsigma.ps} 
\end{figure}

We determined the rotational velocities and turbulent velocities
that belong to the individual line-emitting regions
in the same way as we have done it before for other Seyfert galaxies
(Kollatschny
 \& Zetzl\citealt{kollatschny11,kollatschny13a,kollatschny13b,kollatschny13c}):
Based on the observed linewidths (FWHM) and linewidth ratios
FWHM/$\sigma_{\mathrm{line}}$ we determined
the locations of the individual lines in Fig.~\ref{fwhmsigma.ps}.
In this figure, the grid, resulting from model calculations, presents 
theoretical linewidth ratios based on Lorentzian profiles that
are broadened owing to rotation.
We read the widths of the Lorentzian profiles and
the rotational velocities of the individual lines from their positions between 
the contour lines of constant Lorentzian linewidth and the vertical 
contour lines representing different $v_{\text{rot}}$.
The FWHM/$\sigma_{\mathrm{line}}$ versus FWHM grid based on our
model calculations is publicly 
available.\footnote{http://www.astro.physik.uni-goettingen.de/\textasciitilde{}zetzl/blrvelo/}
The FWHM and FWHM/ $\sigma_{\mathrm{line}}$ values
we obtained for  HE\,1136-2304 are given in
Tab.~\ref{acc_str_h/r.ps} together with the derived $v_{\text{turb}}$ and
$v_\text{rot}$ velocities of the Balmer and Helium lines.
It has been shown that the region of each emission line has a 
characteristic mean turbulent velocity within a narrow range
(Kollatschny \& Zetzl\citealt{kollatschny11,kollatschny13a,kollatschny14}).
We derived the following mean turbulent velocities
belonging to the emitting regions of the individual lines: 
400 \kms for H$\beta$,  
700 \kms for H$\alpha$,
800 \kms for \ion{He}{i}\,$\lambda 5876$, and 
900 \kms for \ion{He}{ii}\,$\lambda 4686$.
 
In the next step,
we determine the heights of the line-emitting regions above the midplane 
 as we have before for other Seyfert galaxies.
The ratio of the turbulent velocity $v_{\text{turb}}$ with respect to the
rotational velocity $v_\text{rot}$ in the line-emitting region
gives us information on the ratio of the
height  $H$ with respect to the radius $R$
 of the line-emitting regions
as presented in Kollatschny \& Zetzl \cite{kollatschny11,kollatschny13a}, i.e.,
\begin{equation}
\label{eq:HtoR}
  H/R = (1/\alpha) (v_{\text{turb}}/v_{\text{rot}}). 
\end{equation}
The unknown viscosity parameter $\alpha$ is assumed to be constant
and to have values of 0.1 to 1 (e.g., Frank et al.\citealt{frank03}).
For simplicity we assume a value of 1 in the present investigation. 
The distance of the line-emitting regions of the individual lines is known
from reverberation mapping (section 3.3). 
We present in Tab.~\ref{acc_str_h/r.ps}
the derived heights above the midplane of
the line-emitting regions in HE\,1136-2304
(in units of light days) and the
ratio $H/R$ for the individual emission lines.

%
\begin{table*}
    \centering
       \leavevmode
       \tabcolsep1.5mm 
        \newcolumntype{d}{D{.}{.}{-2}} 
        \newcolumntype{p}{D{+}{\,\pm\,}{-1}}
        \newcolumntype{K}{D{,}{}{-2}}
\caption{
Line profile parameters and radius and height of
the line-emitting regions
for individual emission lines in HE\,1136-2304.
For the values of $v_{\text{turb,exp}}$, see  Kollatschny et al.\cite{kollatschny13a}.
}

\begin{tabular}{lcKKKKKKlKdc}
 \htopline
\hspace{3mm} Line & \mcc{FWHM} & \mcr{FWHM/$\sigma$} &\mcc{$v_{\text{turb}}$}&\mcc{$v_{\text{turb,exp}}$} &\mcc{$v_{\text{rot}}$} &\mcc{Radius}
                  & \mcc{Height} & \mcr{$H/R$} & \mcc{Height$_{\text{exp}}$} & \mcc{$H_{\text{exp}}/R$}\\
\hspace{3mm}      &\mcc{[\kms{}]}&&\mcc{[\kms{}]}&\mcc{[\kms{}]} &\mcc{[ld]}
                  & \mcc{[ld]} & \mcc{[ld]} & \mcc{} & \mcc{[ld]}\\
\hmidline   
\noalign{\smallskip}
\Ha              &2668&1.47,\pm{0.15}&712,^{+180}_{-164} &700&1528,^{+111}_{-129}&15.0,^{+4.2}_{-3.8}&7.0,\pm{2.7}&0.47&6.9,\pm{2.7}&0.46\\
\Hb              &3791&2.15,\pm{0.20}&492,^{+176}_{-160} &400&2219,^{+89}_{-98}&7.5,^{+4.6}_{-5.7}&1.7,\pm{1.4}&0.23&1.4,\pm{1.2}&0.19\\
\Nl{He}{i}{5876} &4131&1.97,\pm{0.42}&777,^{+619}_{-485} &800&2405,^{+244}_{-350}&7.3,^{+2.8}_{-4.4}&2.4,\pm{2.4}&0.33&2.4,\pm{2.4}&0.33\\
\Nl{He}{ii}{4686}&5328&1.80,\pm{0.35}&1791,^{+1316}_{-960} &900&2983,^{+427}_{-809}&3.0,^{+5.3}_{-3.7}&1.8,\pm{3.5}&0.60&0.9,\pm{2.1}&0.30\\
\noalign{\smallskip}
\hbotline  
\end{tabular}
\label{acc_str_h/r.ps}
\end{table*}

\begin{figure}[t]
\centering
\includegraphics[width=7.6cm,angle=0]{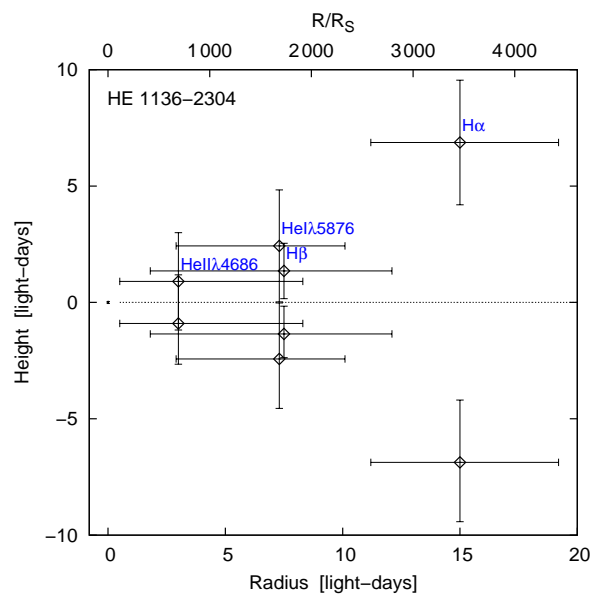} 
 \caption{Structure of the BLR in HE\,1136-2304.
   The dot at radius zero has the size of a Schwarzschild black hole
 (for $M_\text{BH}=3.8\times10^{7}M_{\sun}$) multiplied by a factor of ten.
 }
   \label{disc_he1136.ps}
\end{figure}

The BLR structure of HE\,1136-2304 is shown in Fig.~\ref{disc_he1136.ps}
as a function of
distance to the center and height above the midplane. 
The dot at radius zero gives the size of the Schwarzschild radius
$R_{\mathrm{S}} =4.31\times10^{-3}\,\mathrm{ld} = 1.1\times10^{13}\,\mathrm{cm}$
for a black hole mass
(with $M=3.8.\times10^{7}M_{\sun}$)
 multiplied by a factor of ten.
The label on top of the figure gives the distances of the line-emitting regions
in units of the Schwarzschild radius.
As has been observed previously in other galaxies,
the \ion{He}{ii}\,$\lambda 4686$ 
line originates at the shortest distance from the center
and the smallest height above
the midplane in comparison to the Balmer and HeI lines.
In comparison to H$\beta$, H$\alpha$ originates at a larger distance from
the midplane.

\section{Discussion}

\subsection {Optical variability}

We thoroughly investigated the spectroscopic variability behavior
of HE\,1136-2304 by taking 16 spectra over a period of six months
between February to August 2015. The fractional variability F$_{\text{var}}$
was on the order of 0.1 in the optical
continuum without correcting for the host galaxy flux. After correcting for
the host galaxy contribution, the fractional variability F$_{\text{var}}$
of the continuum amounted to $0.25-0.3$ (see Paper I). 
The integrated Balmer and Helium lines showed F$_{\text{var}}$ values 
 of 0.1 to 0.5 and the higher ionized lines originating closer to the center
varied with stronger amplitudes.
These results describing the continuum and emission line 
variability are similar
to those detected in other variable Seyfert galaxies such as NGC\,5548
(Peterson et al.\citealt{peterson04}),
Mrk\,110  (Kollatschny et al.\citealt{kollatschny01}),
or 3C\,120 (Kollatschny et al.\citealt{kollatschny14}).
This confirms that the variability behavior of this changing look AGN
is similar to that of other Seyfert galaxies.

\subsection {Balmer decrement  variability}

The Balmer decrement \Ha{}/\Hb{} of the narrow components has a value of
2.81. This corresponds exactly to the expected theoretical line ratio
(Case B) without any reddening. However, the Balmer decrement \Ha{}/\Hb{} 
of the broad components varies with the continuum and/or Balmer line intensity.
For example, the broad line Seyfert galaxy NGC~7693 showed the same behavior
based on long-term variability studies over a period of 20 years
(Kollatschny et al.\citealt{kollatschny00}): the Balmer decrement
decreased with increasing H$\beta$ flux.

Heard \& Gaskell\cite{heard16} proposed a model
\text{with} additional
dust reddening clouds interior to the narrow-line region causing higher 
Balmer decrements in the BLR.  
In contrast to this model, there might be important 
optical depth effects in the BLR itself explaining 
the observations.
This is consistent with the finding that H$\alpha$ originates
at twice the distance of H$\beta$.
A similar radial stratification as seen in HE\,1136-2304 has been observed
in, for example, Arp~151 (Bentz et al.\citealt{bentz10}) as well.
It has been discussed by Korista and 
Goad\cite{korista04} that the radial stratification is a result of
optical-depth effects of the Balmer lines: the broad-line
Balmer decrement decreases in high continuum states and
steepens in low states exactly as observed in
HE\,1136-2304 (see Fig.~\ref{balmerdek_vs_cont4570.ps}).
The continuum varied by a factor of nearly two during our campaign
in 2015. However, we did not detect simultaneous variations of the Seyfert 
subtype during our observing period of seven months.
 A variation of Seyfert subtypes might
be connected with stronger continuum amplitudes
and/or longer timescales as has been seen before, for example,
in Fairall\,9 (Kollatschny et al.\citealt{kollatschny85}).

\subsection {H$\beta$ lag versus optical continuum luminosity}

Now we want to test whether HE\,1136-2304
follows the general trend in the radius-luminosity relationship for AGN 
(Kaspi et al.\citealt{kaspi00}; Bentz et al.\citealt{bentz13}).
We determined a continuum luminosity $\log \lambda \text{L}_{\lambda}$
of 42.6054 erg\,s$^{-1}$ ($0.47 \times
10^{-15}$\,erg\,s$^{-1}$\,cm$^{-2}$\,\AA$^{-1}$)
in the optical at 5100\,\AA{} after correction for the contribution of the
host galaxy (Zetzl et al.\citealt{zetzl18}).  
Furthermore, we derived a mean radius of 7.5 light days for the H$\beta$
line-emitting region based on the delay of the integrated
H$\beta$ line variability curve with respect to the optical continuum light
curve. 

\begin{figure}
\centering
\includegraphics[height=9cm,angle=-90]{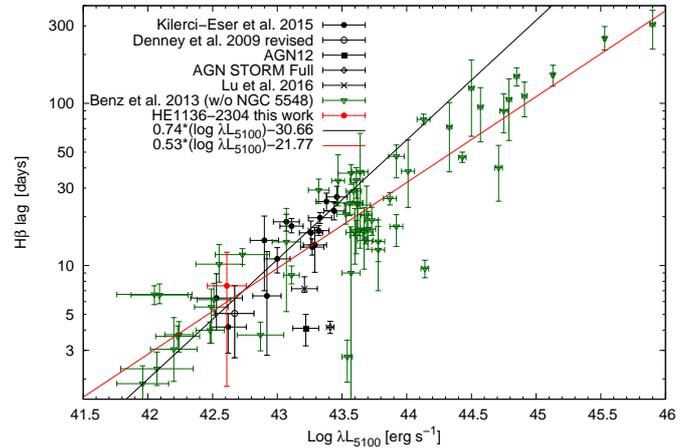}
\caption{Optical continuum luminosity and H$\beta$-optical lags
for HE\,1136-2304 and other AGN.}
\label{hblag_vs_loglambda_L_lambda_v2.ps}
\end{figure}

Fig.~\ref{hblag_vs_loglambda_L_lambda_v2.ps} shows the
optical continuum luminosity and H$\beta$-optical lags
for HE\,1136-2304 (red), for NGC\,5548 based on different variability campaigns
(black; Pei et al.\citealt{pei17} based on
Kilerci-Eser et al.\citealt{kilerci15} and Denney et al.\citealt{denney09}),
and a sample of other AGN excluding NGC\,5548
 (green; Bentz et al.\citealt{bentz13}).
The black line is the linear least-squares fit to the NGC\,5548 data
as presented by Pei et al.\cite{pei17}. The $R-L$(5100\AA{}) relationship is given by
\begin{equation}
 \log \Bigg[\frac{R_\text{BLR}}{ 1\,\text{light-day}}\Bigg] = K +\beta \log\Bigg[\frac{\lambda L_{\lambda}(5100\AA)}{{10^{44}\, \text{erg\, s}}^{-1}}\Bigg],
\end{equation}
where K is the origin and $\beta$ is the slope. 
The red solid line gives the best-fit linear regression to the whole
AGN sample.
The data of HE\,1136-2304 is in very good accordance with the general 
H$\beta$ lag versus the optical continuum luminosity relation.
The red solid line has a slope $\beta$ of 0.53, which is therefore 
identical to the best-fit slope of $0.533^{+0.035}_{-0.033}$
of Bentz et al.\cite{bentz13}. This value is very close to the value
of 0.5 expected from simple photoionization arguments, i.e.,
\begin{equation}
 R \sim L^{1/2}
\end{equation}
(e.g., Kaspi et al.\citealt{kaspi00}; Bentz et al.\citealt{bentz13}
and references therein).

We tested whether the $\beta$ slope approaches values even closer 
to 0.5 or whether the Pearson correlation coefficient becomes
higher if we add a few light
days to the H$\beta$ radius. Such an additional delay
might be caused by the
fact that the optical continuum is generally delayed
by a few light days with respect to the
driving X-ray light curve (Zetzl et al.\citealt{zetzl18},
Shappee et al.\citealt{shappee14}; Fausnaugh et al.\citealt{fausnaugh16}).

\begin{figure}
\centering
\includegraphics[height=9cm,angle=-90]{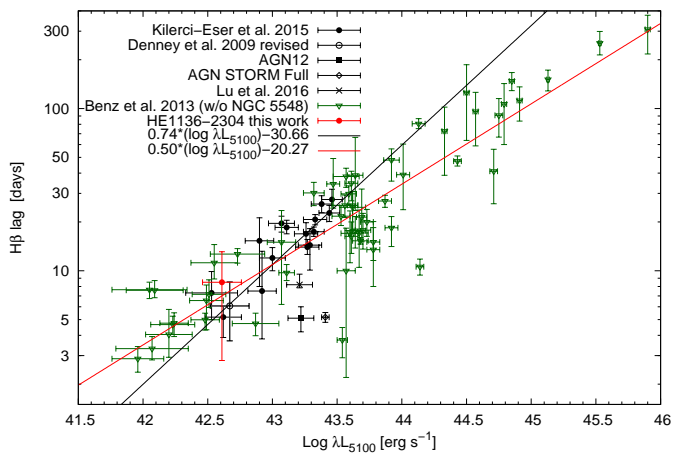}
\caption{Optical continuum luminosity and H$\beta$-optical lags
for HE\,1136-2304 and other AGN (plus 1 day additional lag for optical-X-ray lag).}
\label{hblag_vs_loglambda_L_lambda_v2_plus1days.ps}
\end{figure}

\begin{figure}
\centering
\includegraphics[height=9cm,angle=-90]{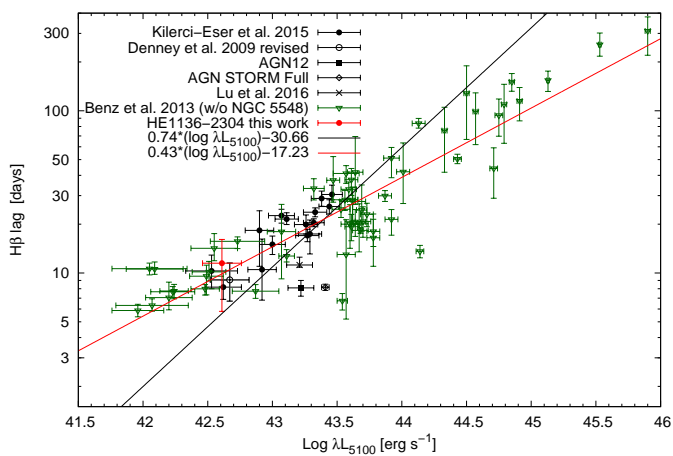}
\caption{Optical continuum luminosity and H$\beta$-optical lags
for HE\,1136-2304 and other AGN (plus 4 days additional lag for optical-X-ray lag).}
\label{hblag_vs_loglambda_L_lambda_v2_plus4days.ps}
\end{figure}

We added additional lags of 
one to eight light days to all  H$\beta$-optical lags to take into account a 
systematic delay of the optical bands
with respect to the driving X-ray flux.  
 A time delay of eight light days is an upper limit based on the correlation
 of the optical band light curves with respect to
the XRT light curve (Zetzl et al.\citealt{zetzl18}).
Figs.~\ref{hblag_vs_loglambda_L_lambda_v2_plus1days.ps} and
\ref{hblag_vs_loglambda_L_lambda_v2_plus4days.ps} show the 
H$\beta$ lag versus optical continuum luminosity diagrams with an additional
lag of one and four days, respectively, taking into account
the optical-X-ray lag.
\begin{table}
    \centering
       \leavevmode
       \tabcolsep5mm 
        \newcolumntype{d}{D{.}{.}{-2}} 
        \newcolumntype{p}{D{+}{\,\pm\,}{-1}}
        \newcolumntype{K}{D{,}{}{-2}}
\caption{ Pearson correlation coefficient for the relation between
 optical continuum luminosities and H$\beta$-optical lags. The H$\beta$ lags
 have been modified assuming additional lags (in units of days)
 for the optical lag with respect to the driving X-ray source. Additionally, we give
the gradient $\beta$.}
\begin{tabular}{cdK}
 \htopline
\hspace{0mm}Offset delay  & \mcc{Pearson CC}  & \mcc{$\beta$}\\
\hspace{0mm}  [days] &                   &              \\
\hmidline   
\noalign{\smallskip}
0             &  0.8870          & 0.529,\pm0.032\\
1             &  0.8903          & 0.496,\pm0.029\\ 
2             &  0.8919          & 0.469,\pm0.027\\
3             &  0.8926          & 0.447,\pm0.026\\
4             &  0.8928          & 0.428,\pm0.025\\ 
5             &  0.8927          & 0.411,\pm0.024\\ 
6             &  0.8923          & 0.396,\pm0.023\\
7             &  0.8918          & 0.383,\pm0.022\\
8             &  0.8911          & 0.371,\pm0.022\\
\noalign{\smallskip}
\hbotline  
\end{tabular}
\label{pearsoncc_lag}
\end{table}
Tab.~\ref{pearsoncc_lag} gives the Pearson correlation coefficient
for the relation between
optical continuum luminosities and H$\beta$-optical lags. The H$\beta$ lags
have been modified assuming additional lags (in units of days)
for the optical lag with respect to the driving X-ray source.
Furthermore, we present the $\beta$ values for the additional
delays that have been assumed.
We get the highest correlation coefficient for an additional delay
of four days. We reached a $\beta$ slope of exactly 0.5 for an additional
delay of one day. \\

\subsection{Structure and kinematics in the BLR}

\subsubsection{Mean and rms line profiles}

The mean and rms line profiles of the broad emission lines give us
information about the kinematics and structure of the line-emitting
BLR region.
Differences in the broad-line widths of the rms and mean profiles
(see Figs.~\ref{velo_meanrms_ha.ps} to \ref{velo_meanrms_he1.ps})
might be caused by a radial stratification of optical depth effects
in these lines  (Korista and Goad\citealt{korista04}).
 Especially the rms profiles of the Balmer lines
in HE\,1136-2304 show
an asymmetric triple structure. Aside from a central component there
were additional blue and red components at $+/-$1\,400~\kms{} 
(see Fig.~\ref{velo_rms_hahb.ps}).
These components are barely visible in the mean profiles.
The additional component in the red wing is by far stronger than
that in the blue wing.
An additional weak blue component, which is nearly symmetrical to the
red component, is apparent in the
rms profile of H$\beta$ (Fig.~\ref{velo_rms_hahb.ps}).
Furthermore, this red rms component varies relatively
stronger in the H$\beta$ line 
than in H$\alpha$.
The additional blue and red components in the line profiles -- in addition
to the central component -- are an indication
that the line-emitting region is connected to the accretion disk.
Such double-peaked profiles are considered to be ubiquitous signatures
of accretion disks
(e.g., Eracleous \& Halpern\citealt{eracleous03}; Gezari et al.\citealt{gezari07}; Shapovalova\citealt{shapovalova13}; 
Storchi-Bergmann et al.\citealt{storchibergmann17}, and references therein).
In some cases these double-peaked profiles become only visible
in the  rms line profiles as in NGC4593, for example
(Kollatschny \& Dietrich\citealt{kollatschny97}).
The variable Seyfert galaxy Akn~120
is another example of a very strong red component showing up
in the H$\beta$ wing within one year
(Kollatschny et al.\citealt{kollatschny81}).

Similar to the line profiles in NGC4593 
(Kollatschny \& Dietrich\citealt{kollatschny97}), the rms line profiles of 
H$\alpha$ and H$\beta$ in HE\,1136-2304 show a steeper red wing and a flatter 
outer blue wing indicating
an additional outflow component (see Fig.~\ref{velo_rms_hahb.ps}).
The outer blue wing is even more pronounced in the higher ionized
Helium lines in comparison to the Balmer lines (see Fig.~\ref{velo_rms_he.ps}),
indicating a stronger outflow in the inner BLR.

\subsubsection{Velocity delay maps}

The 2D-CCFs or velocity-delay maps  shown in Figs.~\ref{ccf2d_ha.ps}
to \ref{ccf2d_he2.ps} contain additional
 information about the structure and kinematics
of the BLR.
We compare the derived velocity delay maps of HE\,1136-2304
with theoretical
models for the structure and kinematics of the BLR
(Welsh et al.\citealt{welsh91};
Horne et al.\citealt{horne04};
Goad et al.\citealt{goad12};  
Grier et al.\citealt{grier13}) 
and with velocity delay maps of other AGN.
All the velocity delay maps are very symmetric
with respect to their line centers at $v=0$ \kms{}. 
The delays in the wings are by far shorter than in the line center.
Such behavior is typical for thin Keplerian disk BLR models 
(Welsh et al.\citealt{welsh91};
Horne et al.\citealt{horne04}; Grier et al.\citealt{grier13}).
There is an indication in the velocity delay maps of the Balmer lines
that the response in the red wing (at v = 3000 to 5000\,\kms)
 is slightly stronger and that it
shows a shorter delay than in the blue wing.  
This might be caused by an additional inflow component
(Welsh et al.\citealt{welsh91}), by hydro-magnetically driven wind
(Horne et al.\citealt{horne04}), or by an additional turbulent component
(Goad et al.\citealt{goad12}).  

The velocity delay maps of other Seyfert galaxies in general show
two different trends: a more symmetrical velocity delay map that is typical
for Keplerian disks or a velocity delay map showing a strong red component 
caused by strong inflow or hydro-magnetically driven wind, and a combination 
of both.
NGC\,4593 (Kollatschny et al.\citealt{kollatschny97}),
3C\,120 (Kollatschny et al.\citealt{kollatschny14}), 
Mrk\,50 (Barth et al.\citealt{barth11}), 
and NGC\,5548 (Pei et al.\citealt{pei17} and references therein)
 show a more
symmetrical velocity delay map.
NGC\,3516 (Denney et al.\citealt{denney10}),
Mrk1501, PG\,2130+099 (Grier et al.\citealt{grier13}) and Mrk\,335
(Du et al.\citealt{du16}) show a dominant
red component.
Velocity delay maps of other galaxies indicate a combination
of dominant Keplerian motion and an additional red component, such as
 Mrk\,110 (Kollatschny et al.\citealt{kollatschny01}) and Arp\,151
(Bentz et al.\citealt{bentz10}).
However, there are three exceptions (Mrk\,817, NGC\,3227, and Mrk\,142)
in which 
only a strong blue component is present in the  velocity delay maps
(Denney et al.\citealt{denney10}; Du et al.\citealt{du16}).
The velocity delay map of the changing look AGN HE\,1136-2304
is similar to that of most other AGN. It shows a dominant
Keplerian motion component with a slightly more intense red component.

\subsection{Vertical BLR structure in a sample of AGN}

The higher ionized broad emission lines originate at smaller radii
as  shown in section 3.3. Furthermore, the integrated
H$\alpha$ originates at a distance of
 15 light days and therefore at twice the distance
of H$\beta$
(see Tab.~\ref{CCF_1D}).
Moreover, it has been shown that the higher ionized lines originate closer
to the midplane of the accretion disk in comparison to the lower ionized lines.
We presented the BLR structure
as a function of distance to the center and height above the midplane 
(Fig.~\ref{disc_he1136.ps}). 
The \ion{He}{ii}\,$\lambda 4686$ line originates
closest to the midplane.
H$\alpha$ originates at a larger distance from
the midplane in comparison to H$\beta$. Such a trend 
has been observed before in other galaxies as NGC~7469 
(Kollatschny \& Zetzl\citealt{kollatschny13c}) and 3C~120
(Kollatschny et al.\citealt{kollatschny14}).

A second trend has been found when comparing 
the H$\beta$ distances above the midplane
for different active galaxies: galaxies showing 
the broadest H$\beta$ linewidths originate closest to the midplane,
while galaxies showing 
the narrowest H$\beta$ linewidths originate at the largest
distance to the midplane
(Kollatschny et al.\citealt{kollatschny14}).
The linewidths (with respect to the individual lines)
are therefore a characteristic for the height of the line-emitting 
regions above the midplane.
We present the
height-to-radius ratio and FWHM of H$\beta$ 
for a sample of AGN (Kollatschny et al.\citealt{kollatschny14})
and for HE\,1136-2304 in Tab.~\ref{height_to_radius}.
\begin{table}
    \centering
       \leavevmode
       \tabcolsep7mm 
        \newcolumntype{d}{D{.}{.}{-2}} 
        \newcolumntype{p}{D{+}{\,\pm\,}{-1}}
        \newcolumntype{K}{D{,}{}{-2}}
\caption{ 
Height-to-radius ratio and FWHM of H$\beta$ 
for a sample of AGN.
}
\begin{tabular}{lKK}
 \htopline
\hspace{3mm} Campaign & \mcc{FWHM}   & \mcc{$H_{\text{obs}}/R$}\\
\hspace{3mm}          &\mcc{[\kms{}]}&                      \\
\hmidline   
\noalign{\smallskip}
NGC 7469      &2169,^{+459}_{-459}&0.36,^{+0.14}_{-0.14}\\
3C 120 p04    &2205,^{+185}_{-185}&0.19,^{+0.03}_{-0.03}\\
3C 120 g12    &2539,^{+466}_{-466}&0.31,^{+0.07}_{-0.07}\\
3C 120 k14    &3252,^{+67}_{-67}&0.27,^{+0.03}_{-0.03}\\
NGC 3783      &3093,^{+529}_{-529}&0.33,^{+0.10}_{-0.10}\\
HE 1136-2304  &3791,^{+150}_{-150}&0.22,^{+0.06}_{-0.06}\\
NGC 5548 T1   &4044,^{+199}_{-199}&0.16,^{+0.03}_{-0.03}\\
NGC 5548 T2   &7202,^{+392}_{-392}&0.25,^{+0.06}_{-0.06}\\
NGC 5548 \Hb{}&5957,^{+224}_{-224}&0.06,^{+0.02}_{-0.02}\\
\dots         &8047,^{+1268}_{-1268}&0.22,^{+0.13}_{-0.13}\\
\dots         &5691,^{+164}_{-164}&0.18,^{+0.03}_{-0.03}\\
\dots         &7202,^{+392}_{-392}&0.25,^{+0.06}_{-0.06}\\
\dots         &6247,^{+343}_{-343}&0.10,^{+0.04}_{-0.04}\\
\dots         &5776,^{+237}_{-237}&0.11,^{+0.03}_{-0.03}\\
\dots         &5706,^{+357}_{-357}&0.10,^{+0.03}_{-0.03}\\
\dots         &5541,^{+354}_{-354}&0.09,^{+0.03}_{-0.03}\\
\dots         &4664,^{+324}_{-324}&0.16,^{+0.04}_{-0.04}\\
\dots         &4044,^{+199}_{-199}&0.16,^{+0.03}_{-0.03}\\
\dots         &6142,^{+289}_{-289}&0.13,^{+0.05}_{-0.05}\\
\dots         &6377,^{+147}_{-147}&0.03,^{+0.01}_{-0.01}\\
\dots         &4596,^{+505}_{-505}&0.12,^{+0.05}_{-0.05}\\
3C 390.3      &9958,^{+1046}_{-1046}&0.06,^{+1.26}_{-1.26}\\
\noalign{\smallskip}
\hbotline  
\end{tabular}
\label{height_to_radius}
\end{table}
\begin{figure}
\centering
\includegraphics[width=6.5cm,angle=-90]{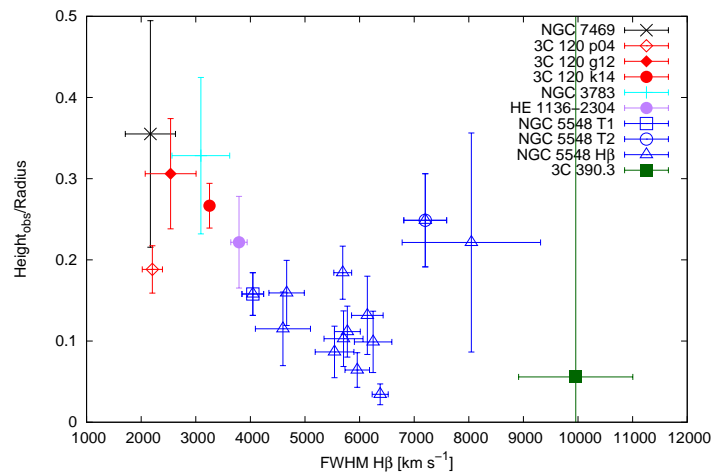}
\caption{Height-to-radius ratio for the H$\beta$ line-emitting regions
for a sample of AGN showing different H$\beta$ linewidths (FWHM).}
\label{korr_h_obs_r_fwhm.ps}
\end{figure}
The height-to-radius ratio for H$\beta$ is largest for galaxies showing
narrow emission lines and smallest for galaxies with broad lines.
The overall picture we derived for the BLR region structure
previously in Kollatschny \& Zetzl\cite{kollatschny13c}
and Kollatschny et al.\citealt{kollatschny14}
 is confirmed 
by the additional emission line data of HE\,1136-2304. The
derived height-to-radius ratio for HE\,1136-2304
confirms the general trend (see Fig.~\ref{korr_h_obs_r_fwhm.ps}).
Again, the  HE\,1136-2304 data support the picture
 that the broad emission line geometries of AGN are not
simply scaled-up versions 
depending only on the central luminosity
(and central black hole mass).

\section{Summary}

We present results of a spectral monitoring campaign
of the changing look AGN HE\,1136-2304 obtained by
 the 10 m SALT telescope between 2014 December and 2015 July.
These observations were taken subsequently to a continuum outburst
detected in the X-rays and in the optical in 2014 July. Our findings
can be summarized as follows:

\begin{enumerate}[(1)]
\item The BLR in HE\,1136-2304 is stratified with respect to the distance of
the individual line-emitting regions. The integrated emission line intensities
 of H$\alpha$, H$\beta$,
\ion{He}{i}\,$\lambda 5876$, and \ion{He}{ii}\,$\lambda 4686$ originate at
distances of $15.0^{+4.2}_{-3.8}$, $7.5^{+4.6}_{-5.7}$,
$7.3^{+2.8}_{-4.4}$, and $3.0^{+5.3}_{-3.7}$
 light days with respect to the optical continuum
at 4570\,\AA{}. The variability amplitudes of the integrated emission lines
are a function of distance to the ionizing source as well.

\item We derived a central black hole mass of
$3.8  \times 10^{7} M_{\odot}$ based on the linewidths, corrected for
the turbulent component, and distances of the
line-emitting regions.

\item Based on velocity delay maps, the light curves of the emission 
line centers are delayed by 10 to 20 days with respect
to the continuum variations.
 The outer line wings of the emission lines respond much faster
to the continuum variations in all lines
indicating an Keplerian
disk component for the broad line-emitting region. The response in
the outer wings is even shorter than the response of the adjacent
optical continuum
flux with respect to the ionizing continuum flux by about two light days. 

\item The vertical BLR structure in HE\,1136-2304
confirms the general trend that line emitting regions in 
 AGN showing narrower emission lines originate at larger
distances from the midplane in comparison to AGN showing broader
emission lines.

\item In general, the variability behavior of the changing look AGN 
HE\,1136-2304 is similar to that of other AGN.

\end{enumerate}

\begin{acknowledgements}
This work has been supported by the DFG grants Ko 857/33-1
and Ha3555/12-2.
\end{acknowledgements}

\clearpage

\begin{appendix}
\section{\bf Additional figures}
\begin{figure*}
\vspace{1cm}\hspace{1cm}
\includegraphics[width=16cm,angle=0]{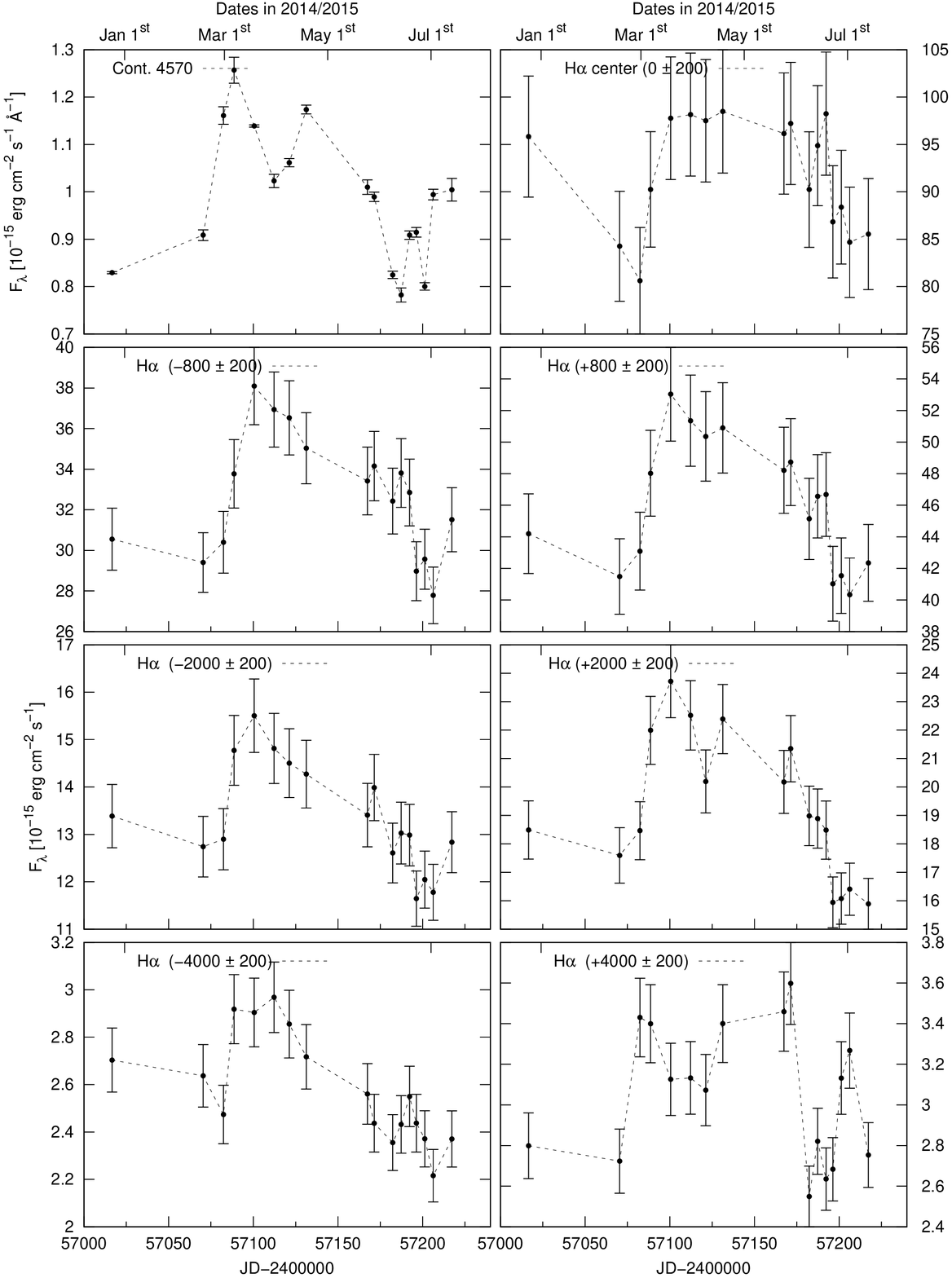}
\vspace{1cm}
      \caption{Light curves of the continuum flux at 4570\,\AA\
  and of selected H$\alpha$ line segments
   (in units of  10$^{-15}$ erg s$^{-1}$ cm$^{-2}$):  H$\alpha_\text{center}$ and segments at
v = $+/-$800, $+/-$2~000, $+/-$4~000 \kms{}. 
}
\label{lc_seg_ha_select.ps}
   \end{figure*}
%
%
   \begin{figure*}
\vspace{1cm}\hspace{1cm}\includegraphics[width=16cm,angle=0]{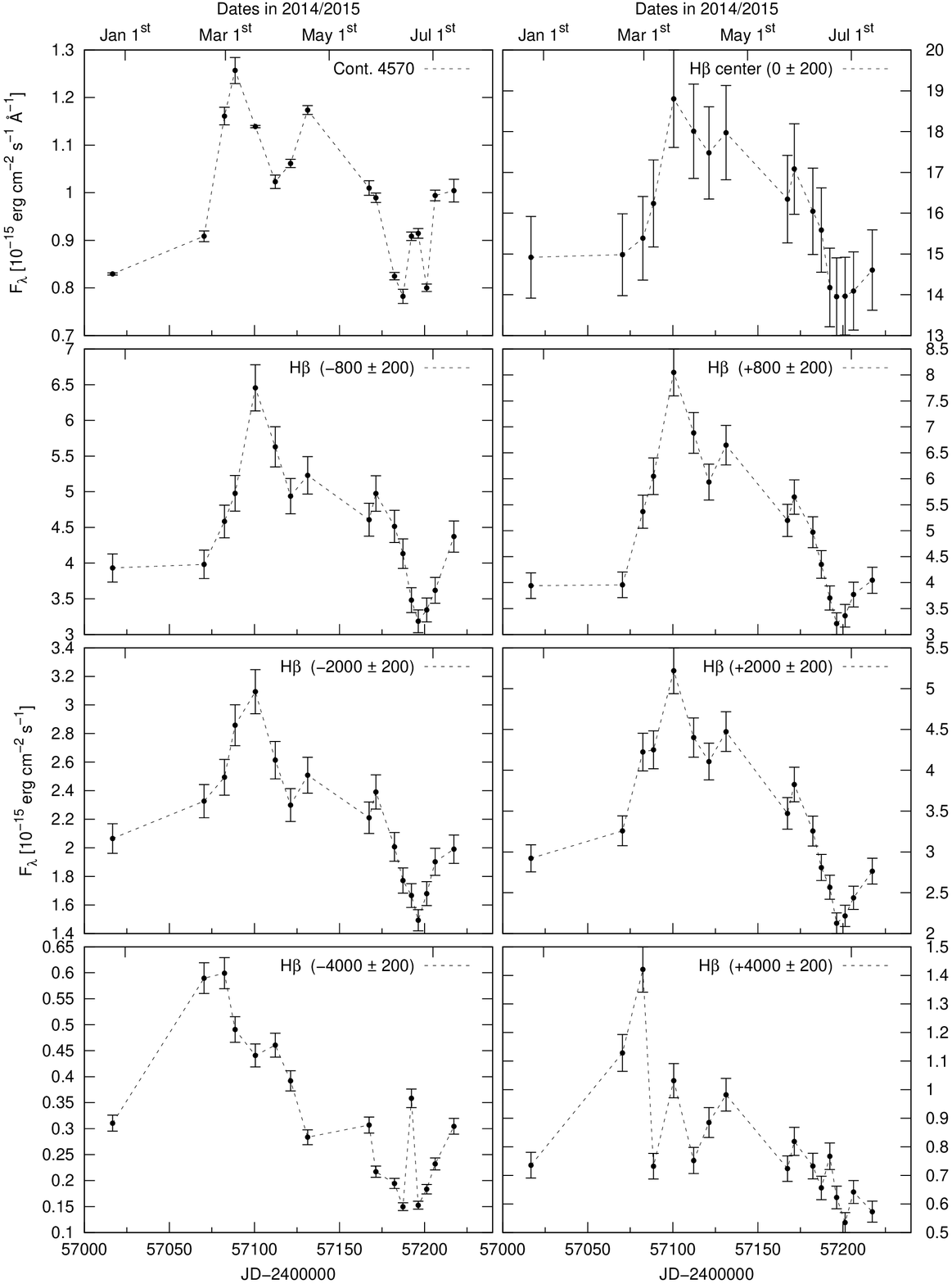}\vspace{1cm}
      \caption{Light curves of the continuum flux at 4570\,\AA\
  and of selected H$\beta$  line segments
   (in units of  10$^{-15}$ erg s$^{-1}$ cm$^{-2}$):  H$\beta_\text{center}$ and segments at
v = $+/-$800, $+/-$2~000, $+/-$4~000 \kms{}. 
}
\label{lc_seg_hb_select.ps}
   \end{figure*}
%
%
   \begin{figure*}
\vspace{1cm}\hspace{1cm}\includegraphics[width=16cm,angle=0]{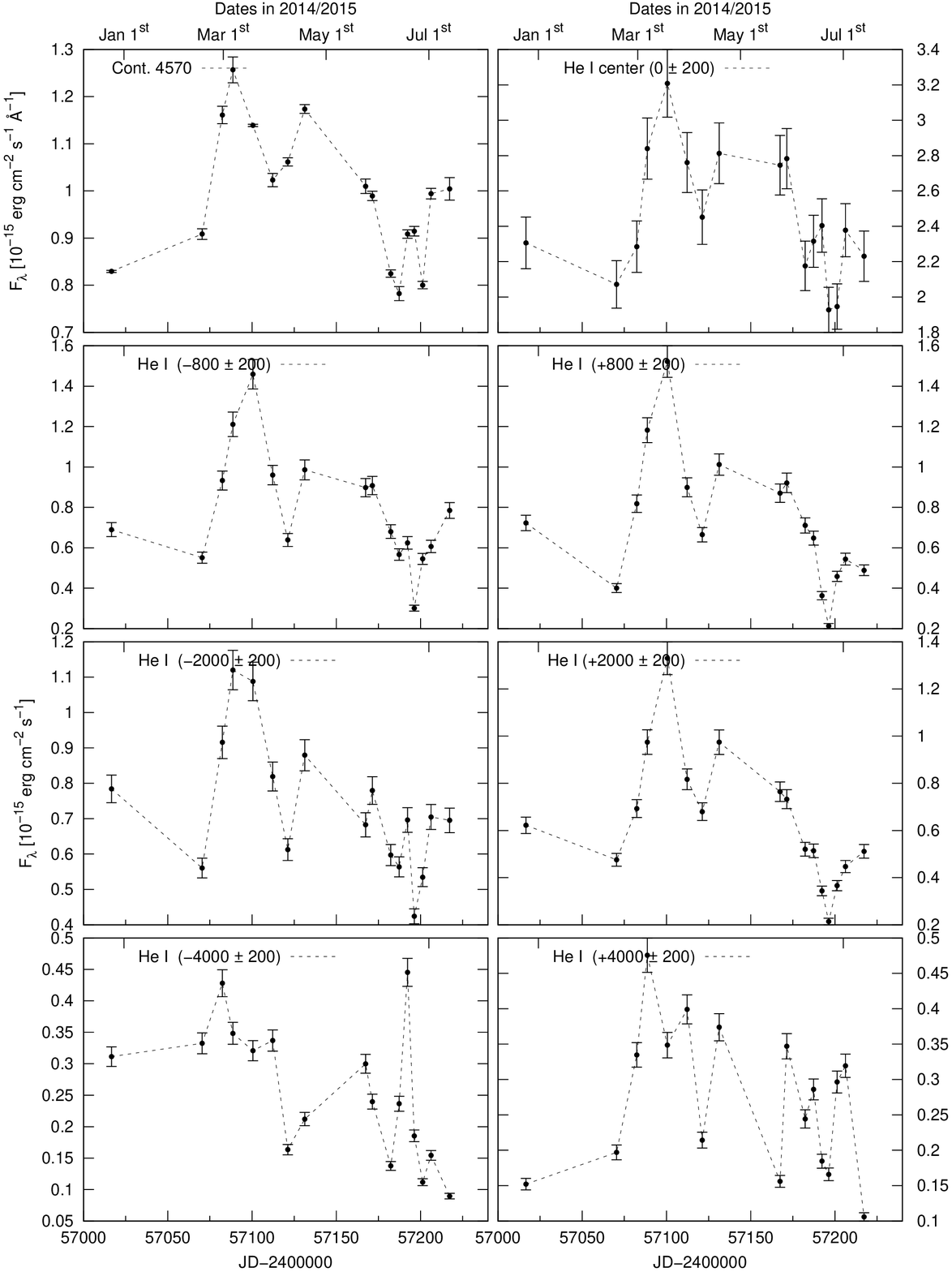}\vspace{1cm}
      \caption{Light curves of the continuum flux at 4570\,\AA\
  and of selected \ion{He}{i}\,$\lambda 5876$  line segments
   (in units of  10$^{-15}$ erg s$^{-1}$ cm$^{-2}$): \ion{He}{i}\,$\lambda 5876_\text{center}$
 and segments at
v = $+/-$800, $+/-$2~000, $+/-$4~000 \kms{}. 
}
\label{lc_seg_he1_select.ps}
   \end{figure*}
%
%
   \begin{figure*}
\vspace{1cm}\hspace{1cm}\includegraphics[width=16cm,angle=0]{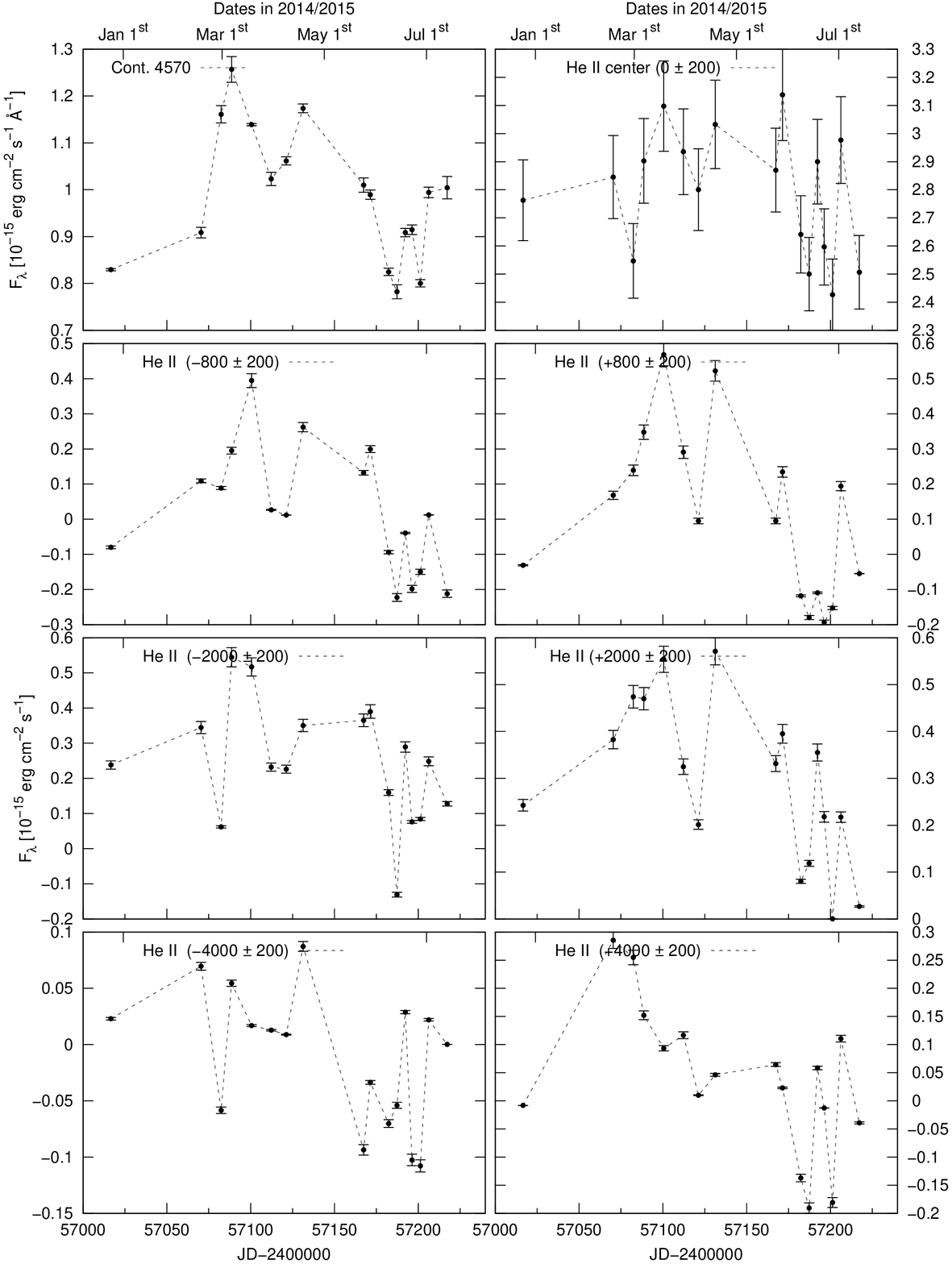}\vspace{1cm}
      \caption{Light curves of the continuum flux at 4570\,\AA\
  and of selected \ion{He}{ii}\,$\lambda 4686$ line segments
   (in units of  10$^{-15}$ erg s$^{-1}$ cm$^{-2}$): \ion{He}{ii}\,$\lambda 4686_\text{center}$
 and segments at
v = $+/-$800, $+/-$2~000, $+/-$4~000 \kms{}. 
}
\label{lc_seg_he2_select.ps}
   \end{figure*}

\end{appendix}


\begin{thebibliography}{}
%
 \bibitem[\protect\citeauthoryear{} {2011}]{barth11} Barth, A.~J., Pancoast, A., Thorman, S.~J.,
 et al. 2011, ApJL, 743, L4
%
 \bibitem[\protect\citeauthoryear{} {2010}]{bentz10} Bentz, M.~C.,
Horne, K., Barth, A.~J., et al. 2010, ApJ, 720, L46
%
 \bibitem[\protect\citeauthoryear{} {2013}]{bentz13} Bentz, M.~C., Denney, K.~D., Grier, C.~J., et al. 2013,
 ApJ, 767, 149
%
 \bibitem[\protect\citeauthoryear{} {2012}]{chelouche12} Chelouche, D., 
  \& Daniel, E. 2012,  ApJ, 747, 62
%
 \bibitem[\protect\citeauthoryear{} {1973}]{souffrin73} Collin-Souffrin, S., Alloin, D., \& Andrillat, Y. 1973, A\&A, 22, 343
%
 \bibitem[\protect\citeauthoryear{} {1995}]{dietrich95} Dietrich, M.,
   \& Kollatschny, W. 1995, A\&A, 303, 405
%
 \bibitem[\protect\citeauthoryear{} {2009}]{denney09} Denney, K.~D., Peterson, B.~M., Pogge, R.~W., et al. 2009,
 ApJL, 704, L80
%
 \bibitem[\protect\citeauthoryear{} {2010}]{denney10} Denney, K.~D., Peterson, B.~M., Pogge, R.~W., et al. 2010,
 ApJ, 721, 715
%
\bibitem[\protect\citeauthoryear{} {2014}]{denney14} Denney, K.~D., De Rosa, G., Croxall, K., et al. 2014,
 ApJ, 796, 134
%
\bibitem[\protect\citeauthoryear{} {2016}]{du16} Du, P., Lu, K.~X., Hu, C., et al. 2016,
 ApJ, 820, 27
%
 \bibitem[\protect\citeauthoryear{} {1988}]{edelson88} Edelson, R.,
  \&  Krolik, J. 1988, ApJ, 333, 646
%
 \bibitem[\protect\citeauthoryear{} {2016}]{edelson15} Edelson, R., Gelbord, J.~M., Horne, K., et al. 2015,
 ApJ, 806, 129
%
 \bibitem[\protect\citeauthoryear{} {2003}]{eracleous03} Eracleous, M.,
  \& Halpern, J. 2003, ApJ, 599, 886
%
 \bibitem[\protect\citeauthoryear{} {2016}]{fausnaugh16} Fausnaugh, M.~M., Denney, K.~D., Barth, A.~J., et al.
2016, ApJ, 821, 56
%
\bibitem[\protect\citeauthoryear{} {2003}]{frank03} Frank, J., King, A.,
 \& Raine, D. 2003, Accretion Power in Astrophysics, Cambridge University Press
%
 \bibitem[\protect\citeauthoryear{} {2000}]{fromerth00} Fromerth, M.~J.,
 \& Melia, F. 2000, ApJ, 533, 172 
%
\bibitem[\protect\citeauthoryear{}{2009}]{gaskell09} Gaskell, C.~M. 2009,
 NewAR, 53, 114
%
 \bibitem[\protect\citeauthoryear{} {1987}]{gaskell87} Gaskell, C.~M., \& 
 Peterson, B.~M. 1987, ApJS 65, 1
%
 \bibitem[\protect\citeauthoryear{} {2007}]{gezari07} Gezari, S., Halpern, J.~P., \& Eracleous, M. 2007, ApJS 169, 167
%
 \bibitem[\protect\citeauthoryear{}{2012}]{goad12} Goad, M.~R., Korista, K.~T.,
 \& Ruff, A.~J. 2012, MNRAS, 426, 3086
%
 \bibitem[\protect\citeauthoryear{} {2011}]{graham11} Graham, A.~W., Onken, C.~A., Athanassoula, E., \& Combes, F. 2011, MNRAS 412, 2211
%
 \bibitem[\protect\citeauthoryear{} {2012}]{grier12} Grier, C.~J., Peterson,
B.~M., Pogge, R.~W., et al. 2012, ApJ 755, 60
%
 \bibitem[\protect\citeauthoryear{} {2013}]{grier13} Grier, C.~J., Peterson,
B.~M., Horne, K., et al. 2013, ApJ 764, 47
%
 \bibitem[\protect\citeauthoryear{}{2016}]{heard16} Heard, C.~Z., \&
 Gaskell, C.~M. 2016, MNRAS, 461, 4227
%
\bibitem[\protect\citeauthoryear{} {2004}]{horne04} Horne, K., Peterson, B.~M.
, Collier, S.~J., \& Netzer, H. 2004, PASP 116,465 
%
 \bibitem[\protect\citeauthoryear{} {2000}]{kaspi00} Kaspi, S., Smith, P.~S., Netzer, H., et al.
  2000, ApJ 533, 631
%
 \bibitem[\protect\citeauthoryear{} {2015}]{kilerci15} Kilerci Eser, E., Vestergaard, M., Peterson, B.~M., Denney, K.~D., \& Bentz, M.~C. 2015, ApJ 801, 8
%
 \bibitem[\protect\citeauthoryear{} {2003}]{kollatschny03} Kollatschny, W. 2003, A\&A, 407, 461
%
  \bibitem[\protect\citeauthoryear{} {2000}]{kollatschny00} Kollatschny, W.,
       Bischoff, K., \& Dietrich, M. 2000, A\&A, 361, 901 
%
  \bibitem[\protect\citeauthoryear{} {2001}]{kollatschny01} Kollatschny, W., Bischoff, K., Robinson,
           E. L., Welsh, W.~F., \& Hill, G.~J. 2001, A\&A, 379, 125 
%
  \bibitem[\protect\citeauthoryear{} {2002}]{kollatschny02} Kollatschny, W., \&
 Bischoff, K. 2002, A\&A, 386, L19 
%
  \bibitem[\protect\citeauthoryear{} {1996}]{kollatschny96} Kollatschny, W., \& Dietrich, M. 1996, A\&A, 314, 43
%
  \bibitem[\protect\citeauthoryear{} {1997}]{kollatschny97} Kollatschny, W., \& Dietrich, M. 1997,
       A\&A, 323, 5
%
 \bibitem[\protect\citeauthoryear{} {1981}]{kollatschny81}  Kollatschny, W., Fricke, K.~J., Schleicher, H., \& Yorke, H.~W. 1981, A\&A, 102, L23
%
  \bibitem[\protect\citeauthoryear{} {1985}]{kollatschny85} Kollatschny, W., \& Fricke, K.~J., 1985, A\&A, 146, L11
%
  \bibitem[\protect\citeauthoryear{} {2014}]{kollatschny14} Kollatschny, W.,
 Ulbrich, K., Zetzl, M., Kaspi, S., \& Haas, M. 2014, A\&A, 566, A106
%
 \bibitem[\protect\citeauthoryear{}{2010}]{kollatschny10} Kollatschny, W., \& Zetzl, M. 2010, A\&A, 522, 36
%
   \bibitem[\protect\citeauthoryear{}{2011}]{kollatschny11} Kollatschny, W., \& Zetzl, M. 2011, Nature, 470, 366 
 %
 \bibitem[\protect\citeauthoryear{}{2013a}]{kollatschny13a} Kollatschny, W., \& Zetzl, M. 2013a, A\&A, 549, A100 
%
 \bibitem[\protect\citeauthoryear{}{2013b}]{kollatschny13b} Kollatschny, W., \& Zetzl, M. 2013b, A\&A Letters, 551, L6 
%
 \bibitem[\protect\citeauthoryear{}{2013c}]{kollatschny13c} Kollatschny, W., \& Zetzl, M. 2013c, A\&A, 558, A26
%
 \bibitem[\protect\citeauthoryear{}{2008}]{komossa08} Komossa S., Zhou H., Wang T., et al. 2008, ApJ, 678, L13
%
 \bibitem[\protect\citeauthoryear{} {1991}]{koratkar91} Koratkar, A.~P., \& Gaskell, M. 1991, ApJ 370, L61
%
 \bibitem[\protect\citeauthoryear{} {2004}]{korista04} Korista, K.~T.,\& Goad, M.~R. 2004, ApJ 606, 749
%
\bibitem[\protect\citeauthoryear{}{2015}]{lamassa15} LaMassa, S.~M., Cales, S., Moran, E.~C., et al.
 2015, ApJ, 800, 144
%
  \bibitem[\protect\citeauthoryear{} {2016}]{macleod16} MacLeod, C.~L., Ross, N.~P., Lawrence, A., et al. 2016, MNRAS, 457, 389 
%
  \bibitem[\protect\citeauthoryear{} {2004}]{onken04} Onken, C.~A., Ferrarese, L., Merritt, D., et al. 2004, ApJ, 615, 645
%
  \bibitem[\protect\citeauthoryear{} {1981}]{osterbrock81} Osterbrock, D.~E.
  1981, ApJ, 249, 462
%
  \bibitem[\protect\citeauthoryear{} {2016}]{parker16} Parker, M.~L., Komossa,
 S., Kollatschny, W., et al. 2016, MNRAS, 461, 1927 
%
 \bibitem[\protect\citeauthoryear{} {2017}]{pei17} Pei, L., Fausnaugh, M.~M., Barth, A.~J., et al.
 2017, ApJ, 837, 131
%
 \bibitem[\protect\citeauthoryear{} {1984}]{penston84} Penston, M.~V., \&
Perez, E. 1984, MNRAS, 211, 33 
%
 \bibitem[\protect\citeauthoryear{} {2002}]{peterson02} Peterson, B.~M., Berlind, P., Bertram, R., et al.
 2002, ApJ, 581, 197
%
  \bibitem[\protect\citeauthoryear{} {2004}]{peterson04} Peterson, B.~M., Ferrarese, L., Gilbert, K.~M., et al.
 2004, ApJ, 613, 682
%
  \bibitem[\protect\citeauthoryear{} {1998}]{peterson98} Peterson, B.~M.,
  Wanders, I., Bertram, R., et al. 1998, ApJ, 501, 82
 %
 \bibitem[\protect\citeauthoryear{} {2016}]{runnoe16} Runnoe, J.~C., Cales, S., Ruan, J.~J., et al. 2016
 2016, MNRAS, 455, 1691 
%
 \bibitem[\protect\citeauthoryear{}{2010}]{shapovalova10} Shapovalova, A.~I., Popovi\'{c}, L.~\'{C}., Burenkov, A.~N., et al. 2010, A\&A, 517, 42
%
\bibitem[\protect\citeauthoryear{}{2013}]{shapovalova13} Shapovalova, A.~I., Popovi\'{c}, L.~\'{C}., Burenkov, A.~N., et al. 2013, A\&A, 559, A10
%
  \bibitem[\protect\citeauthoryear{} {2014}]{shappee14} Shappee, B.~J., Prieto, J.~L., Grupe, D., et al.
 2014, ApJ, 788, 48
%
 \bibitem[\protect\citeauthoryear{} {2003}]{storchi03} Storchi-Bergmann, T., Nemmen da Silva, R., Eracleous, M., et al. 2003, ApJ, 598, 956
%
 \bibitem[\protect\citeauthoryear{} {2017}]{storchibergmann17} Storchi-Bergmann, T., Schimoia, J.~S., Peterson, B.~M., et al. 2017, ApJ, 835, 236
%
 \bibitem[\protect\citeauthoryear{}{1996}]{reimers96} Reimers, D., Koehler, T., \& Wisotzki, L. 1996, A\&AS, 115, 235
%
 \bibitem[\protect\citeauthoryear{} {1991}]{welsh91} Welsh, W.~F., \& Horne, K.
  1991, ApJ, 379, 586
%
\bibitem[\protect\citeauthoryear{} {2006}]{wright06} Wright, E.~L., 2006, PASP, 118, 1711
%
 \bibitem[\protect\citeauthoryear{} {2018}]{zetzl18} Zetzl, M., Kollatschny, W., Ochmann, M.~W., et al.\ 2018, A\&A, 618, A83  (Paper 1)
%

\end{thebibliography}
\end{document}